\documentclass[12pt]{article}
\usepackage{amssymb,amsthm,amsfonts}
\usepackage{epsfig}
\DeclareMathAlphabet{\mathpzc}{OT1}{pzc}{m}{it}
\usepackage[T1]{fontenc}
\usepackage[latin2]{inputenc}
\usepackage{colordvi}
\def\bea{\begin{eqnarray}}
\def\eea{\end{eqnarray}}
\def\bq{\begin{quote}}
\def\eq{\end{quote}}

\def\gappeq{\mathrel{\rlap
{\raise.5ex\hbox{$>$}}
{\lower.5ex\hbox{$\sim$}}}}
\def\lappeq{\mathrel{\rlap{\raise.5ex\hbox{$<$}}
{\lower.5ex\hbox{$\sim$}}}}
\def\simlt{\stackrel{<}{{}_\sim}}

\newcommand{\beq}{\begin{equation}}
\newcommand{\eeq}{\end{equation}}

\def\varkappa{\omega}

\newcounter{mnotecount}[section]

\begin{document}
\pagestyle{empty}
\begin{flushright}
IFT-2005/2\\
CERN-PH-TH/2005-011\\
DESY-05-023\\
{\bf \today}
\end{flushright}
\vspace*{9mm}
\begin{center}

{\Large\bf Flavour violation in general supergravity}\\
\vspace*{1.5cm}
{\bf Piotr~H.~Chankowski}$^1$, 
{\bf Oleg Lebedev}$^2$ and {\bf Stefan~Pokorski}$^{1,3}$\\
\vspace{0.5cm}

$^1$ Institute of Theoretical Physics, Warsaw University, 
Ho\.za 69, 00-681, Warsaw, Poland\\
$^2$ DESY, Theory Group, Notkestra$\beta$e 85, 22603 Hamburg, Germany\\
$^3$ Theoretical Physics Division, Physics Department, CERN, CH-1211 
Geneva 23, Switzerland\\

\vspace*{1.7cm} 
{\bf Abstract} 
\end{center}
\vspace*{0mm}
\noindent
{
We reappraise the flavour changing neutral currents (FCNC) problem in 
string--derived supergravity models. We overview and classify possible 
sources of flavour violation and find that the problem often does not 
arise in classes of models which generate hierarchical Yukawa matrices. 
In such models, constraints from the $K-$ and $D-$meson systems 
leave room for substantial flavour non-universality of the soft terms.
The current $B-$physics experiments only begin to probe its natural range. 
Correlations among different observables can allow one to read off the 
chirality structure of flavour violating sources. We briefly discuss the 
lepton sector where the problem of FCNC is indeed serious and perhaps 
points at an additional symmetry or flavour universality.

}
\vspace*{1.0cm}
\date{\today} 


\vspace*{0.2cm}

\vfill\eject

\newpage

\setcounter{page}{1}
\pagestyle{plain}

\section{Introduction}

The purpose of this paper is to classify different sources of flavour 
violation in supergravity (SUGRA) theories and study the FCNC problem 
associated with these sources. Concerning the latter, it is important 
to ask the ``right'' question, that is, how problematic are the FCNC 
in models that successfully generate the  Yukawa matrices rather than 
in some ``generic'' framework. 

The strategy we pursue in this paper is as follows. First, we classify
various sources of flavour violation in general supergravity models.
We concentrate on certain ``benchmark''  textures of the soft terms. Then, 
using a number of representative Yukawa textures, we evolve the soft terms 
to low energies and study implications of the generated flavour changing 
neutral currents. In some cases such as models with flavour violation 
through the K\"ahler potential or democratic Yukawa textures, the problem 
is severe. On the other hand, when flavour structures are due to different 
localizations of matter fields in the compact space, the FCNC are well 
suppressed. This is in fact a common situation in string models which produce 
the Yukawa coupling hierarchy (e.g. heterotic string or intersecting brane 
models). In string theory, the  mechanisms that can  generate the fermion mass 
hierarchy are quite constraining. As a result, the soft supersymmetry 
(SUSY) breaking terms in superstring derived SUGRA models are of special 
forms and the problem of FCNC becomes mild or simply disappears.
A natural consequence of such models is that the mixing between the  second and 
third generation squarks is expected to be substantial and to lead to 
effects potentially detectable at $B-$factories.

We briefly discuss the lepton sector in which a different picture emerges
and the expected flavour violation exceeds the experimental limits. This 
perhaps signals an additional symmetry or flavour universality of the 
soft terms, which will be probed further in the upcoming round of 
$\mu\rightarrow e\gamma$ experiments.

Finally, we discuss how correlations among various $B-$physics 
observables would allow one to pinpoint the source of flavour violation.

\section{Classification of flavour--violating sources in supergravity}

The $N=1$ supergravity Lagrangian is determined by 3 functions: the 
K\"ahler potential $K$, the superpotential $W$ and the gauge kinetic 
function $f$ (see \cite{Nilles:1983ge} for a review). These are 
functions of the hidden and observable sector fields. Since the 
characteristic mass scale of the observable fields $\phi^\alpha$
is much smaller than the Planck scale, one can expand $K$ and $W$ as
\begin{eqnarray}
&&K=\hat K + \tilde K_{\bar\alpha\beta}\phi^{\ast\bar\alpha}\phi^\beta
   + \left({1\over2}Z_{\alpha\beta}\phi^\alpha\phi^\beta
   + {\rm H.c.}\right)+...~,\nonumber\\
&&W= \hat W + {1\over2}\mu_{\alpha\beta}\phi^\alpha\phi^\beta
   + {1\over 6}
     {\cal Y}_{\alpha\beta\gamma}\phi^\alpha\phi^\beta\phi^\gamma 
   +...~,
\end{eqnarray}
where all parameters are generally functions of the hidden sector fields.
Once the  hidden sector fields develop (large) vacuum expectation values,
these parameters will play a role of various couplings in the observable 
sector. Generally, the resulting kinetic terms for the observable sector
fields are neither canonical nor diagonal, so in order to obtain physical
fields, further diagonalization and rescaling are required.

The soft SUSY breaking terms are obtained from the general supergravity
scalar potential by fixing the gravitino mass $m_{3/2}$ and the VEVs of 
the hidden sector fields, while sending the Planck mass to infinity, 
$M_{\rm Pl}\rightarrow\infty$ \cite{Soni:1983rm,Kaplunovsky:1993rd}.
The fermion Yukawa couplings are then rescaled as
\begin{equation}
{\cal Y}_{\alpha\beta\gamma}^\prime={\hat W^\ast\over\vert\hat W\vert}
e^{\hat K/2}{\cal Y}_{\alpha\beta\gamma} \;.
\end{equation}
The resulting soft SUSY breaking potential relevant to flavour physics 
is given by \cite{Brignole:1997dp} 
\begin{eqnarray}
V_{\rm soft}={\mathpzc m}_{\bar\alpha\beta}^2 \phi^{\ast\bar\alpha}  
              \phi^\beta+\biggl({1\over6}{\cal A}_{\alpha\beta\gamma}   
              \phi^\alpha\phi^\beta\phi^\gamma +{\rm H.c.}\biggr)~,
\label{eqn:susybreakingl}
\end{eqnarray}
where 
\begin{eqnarray}
{\mathpzc m}_{\bar\alpha\beta}^2&=&(m_{3/2}^2+V_0)
             \tilde K_{\bar\alpha\beta}- \bar F^{\bar m} F^n 
             (\partial_{\bar m}\partial_n\tilde K_{\bar\alpha\beta}
             -\tilde K^{\gamma\bar\delta}~\partial_{\bar m}
             \tilde K_{\bar\alpha\gamma}  ~
             \partial_{n} \tilde K_{\bar\delta\beta})~,\nonumber\\
{\cal A}_{\alpha\beta\gamma}&=& {\hat W^\ast\over\vert\hat W\vert}
             e^{\hat K/2}F^m \Bigl[\hat K_m {\cal Y}_{\alpha\beta\gamma}
             +\partial_m {\cal Y}_{\alpha\beta\gamma}\nonumber\\
             &-&\Bigl(\tilde K^{\delta\bar\rho}~ 
             \partial_m \tilde K_{\bar\rho\alpha}~
             {\cal Y}_{\delta\beta\gamma} + (\alpha\leftrightarrow\beta)
             + (\alpha\leftrightarrow\gamma)\Bigr)\Bigr]~.
\label{eqn:mA}
\end{eqnarray}
Here $V_0$ is the vacuum energy, $\partial_m$ denotes differentiation 
with respect to the $m$-th hidden sector field 
($\hat K_m\equiv\partial_m\hat K$), $F^m$ are the SUSY breaking 
$F-$terms and $\tilde K^{\gamma\bar\delta}$ is the inverse of the 
K\"ahler metric $\tilde K_{\bar\delta\gamma}$.

Let us now make our notation more transparent. Greek indices 
$\alpha,\dots$ run over all MSSM \footnote{For a review, see 
\cite{Chung:2003fi}.} 
superfields. However, only fields with appropriate quantum numbers can 
couple together. For instance, the K\"ahler mixing is allowed only for fields 
with the same SM quantum numbers, i.e. only intergenerational mixings of 
$Q_i$, $U_i$, $D_i$ ($i$=1,2,3) are permitted. The allowed Yukawa couplings 
in the quark sector are of the type $Q_iD_jH_1$ or $Q_iU_jH_2$. In the 
Yukawa matrices and the $A-$terms, it is convenient to fix the notation  
as follows: the first index is to refer to the quark doublets, the 
second to the quark singlets, and the last to the Higgs field, e.g. 
$Y_{Q_iD_jH_1}\equiv Y_{ij}^d$. 

Physical fields are obtained by diagonalizing and rescaling the K\"ahler
metric responsible for kinetic terms of the observable sector fields. 
The canonically normalized superfields $\varphi^a$ ($a=Q,D,U$) are 
given by
\begin{equation}
\phi^a= H^a\varphi^a
\end{equation}
(no summation over $a$), with 
\begin{equation}
H^a= V^a~{\rm diag}\left({1\over\sqrt{\tilde K^a_1}},
                         {1\over\sqrt{\tilde K^a_2}},
                         {1\over\sqrt{\tilde K^a_3}} \right)~.
\end{equation}
Here $V^a$ are 3$\times$3 unitary matrices diagonalizing the appropriate 
subblocks $\tilde K^a_{\bar\alpha\beta}$ of the K\"ahler metric and 
$\tilde K^a_{1-3}$ are the corresponding eigenvalues,
\begin{equation}
(V^a)^\dagger \tilde K^a V^a = {\rm diag}\left( \tilde K^a_1,
\tilde K^a_2, \tilde K^a_3\right)~.
\end{equation}
The Yukawa couplings and the soft terms transform correspondingly,
such that the seven $physical$  3$\times$3 flavour structures are 
\begin{eqnarray}
Y^u={1\over\sqrt{\tilde K_{H_2}}}(H^q)^T{\cal Y}^u H^u~,\phantom{aaa}
Y^d={1\over\sqrt{\tilde K_{H_1}}}(H^q)^T{\cal Y}^d H^d~,\nonumber\\
A^u={1\over\sqrt{\tilde K_{H_2}}}(H^q)^T{\cal A}^u H^u~,\phantom{aaa}
A^d={1\over\sqrt{\tilde K_{H_1}}}(H^q)^T{\cal A}^d H^d~,
\end{eqnarray}
\begin{eqnarray}
m^2_Q = H^{q\dagger} {\mathpzc m}^2_Q H^q~,\phantom{aa}
m^2_U = H^{u\dagger} {\mathpzc m}^2_U H^u~,\phantom{aa}
m^2_D = H^{d\dagger} {\mathpzc m}^2_D H^d~,\nonumber
\end{eqnarray}
where the factors $1/\sqrt{\tilde K_{H_{1,2}}}$ account for rescaling of
the Higgs fields. These structures are the sources of flavour violation 
in the MSSM. Clearly, flavour dependence in the Yukawa matrices is 
mandatory, whereas the soft terms serve as additional sources of flavour 
violation. The underlying reason for these additional sources is the 
K\"ahler potential and Yukawa couplings dependence on the hidden sector 
fields: indeed, as seen from Eq.~(\ref{eqn:mA}), if the K\"ahler metric 
for observables fields and the Yukawa couplings were mere constants, the 
soft masses would be proportional to the unit matrix in the physical 
basis and the $A-$terms would be proportional to the Yukawa matrix.

Generally, flavour structures of the soft masses and the $A-$terms are 
independent. In particular, unlike the soft masses, the $A-$terms 
receive a contribution from $\partial_m{\cal Y}_{\alpha\beta\gamma}$. 
Thus, even if the K\"ahler potential is trivial, the $A-$term structure 
can be quite rich.

It is also manifest from Eq.~(\ref{eqn:mA}) that a non-trivial K\"ahler 
potential generally induces flavour violation in both the soft masses and
the $A-$terms. Yet, it is conceivable that such flavour violating terms 
may cancel out in the $A-$terms, due to some deeper dynamical reason. 
The soft masses would then be the only source of flavour violation (in 
addition to the Yukawa couplings).

In general, the soft breaking terms violate CP. CP violating phases have 
two sources. Firstly, they are induced by complex SUSY breaking $F-$terms, $F^m$, 
which also generate CP phases in flavour--independent quantities such as 
gaugino masses, the $\mu$-term, etc. 
Secondly, CP phases appear due to complex SUSY preserving quantities such
as the Yukawa couplings and the K\"ahler metric\footnote{The reparametrization 
invariant measures of CP violation are 
given by  quantities of the type 
Arg$({\cal A}_{\alpha\beta\gamma}^\ast{\cal Y}_{\alpha \beta \gamma})$ 
\cite{Lebedev:2002wq}. These can be non-vanishing even if all $F^m$ are real.}. 
Both sources are problematic for phenomenology, which will be discussed 
below.

Clearly, there are many possibilities which have different motivations 
and distinct phenomenology. Below we classify them.

\subsection{Flavour violation through the Yukawa couplings only}

It is possible that flavour dependence does not appear in the soft 
terms \cite{Nath:1983fp}. This occurs, for example, when 
\begin{equation}
F^m \partial_m\tilde K_{\bar\alpha\beta} = 
F^m \bar F^{\bar n}\partial_m \partial_{\bar n} 
\tilde K_{\bar\alpha\beta}=
F^m \partial_m {\cal Y}_{\alpha\beta\gamma} = 0\;.
\end{equation}
This essentially means that the hidden sector fields that generate 
flavour dependence do not break supersymmetry. 

The most common example of this situation is the dilaton dominated SUSY 
breaking scenario \cite{Kaplunovsky:1993rd}. In this case, the only SUSY 
breaking field is the dilaton $S$ which produces no flavour dependence,
\begin{equation} 
F^S \neq0,~ F^m =0 ~{\rm{for}} ~m\neq S \;.
\end{equation}
Then, the soft terms are universal at the string scale and we have a 
version of the minimal supergravity (minimal SUGRA) model with the sfermion 
soft terms parametrized  by $m^2_0$ and $A_0$:
\begin{equation}
(m^2_{Q,U,D})_{ij} = m^2_0 ~\delta_{ij}~,\phantom{aaa}
A^{u,d}_{ij} =  A_0~Y^{u,d}_{ij}~.
\end{equation}

\subsection{Additional flavour violation in $A-$terms only }

If, for example,
\begin{equation}
F^m \partial_m {\cal Y}_{\alpha\beta\gamma}\not\propto
{\cal Y}_{\alpha\beta\gamma} \;,
\label{eqn:A-non}
\end{equation}
but
\begin{equation}
F^m \partial_m \tilde K_{\bar\alpha\beta}\propto 
F^m \bar F^{\bar n}\partial_m \partial_{\bar n}\tilde K_{\bar\alpha\beta}
\propto\tilde K_{\bar\alpha\beta} \;,
\end{equation}
flavour dependence appears only in the $A-$terms \cite{Abel:2001cv}. 
This means that the hidden sector fields responsible for flavour 
dependence of the K\"ahler potential (if present at all) do not break 
supersymmetry and, moreover, are different from those generating flavour 
dependence of the Yukawa couplings.

This is a rather common situation in string models. Indeed, the $A-$terms
are trilinear parameters and are closely related to the Yukawa couplings,
whereas the soft masses are bilinear and more akin to the K\"ahler 
potential. Thus, generally, they are not directly related to each other.

For example, in the heterotic string models the Yukawa hierarchies are 
naturally produced  if the matter fields are twisted 
\cite{Hamidi:1986vh}-\cite{Casas:1989qx}, i.e. localized at special 
points in the compactified space. In such models, the hidden sector 
field that enters the Yukawa coupling is the $T-$modulus, 
${\cal Y}_{\alpha\beta\gamma}={\cal Y}_{\alpha \beta \gamma}(T)$. 
Generally,
\begin{equation} 
F^T\neq0
\end{equation}
and $\partial_m {\cal Y}_{\alpha \beta \gamma}$ are $not$ proportional to 
${\cal Y}_{\alpha\beta\gamma}$, which leads to non-trivial $A-$terms. On 
the other hand, the K\"ahler metric is diagonal for twisted states
and depends on modular weights of these states:
\begin{equation}
\tilde K_{\bar\alpha\beta}
=\delta_{\bar\alpha\beta}(T+\bar T)^{n_{\bar\alpha}} ~, 
\label{eqn:K}
\end{equation}
The modular weights $n_\alpha$ are constrained by the string selection 
rules for the Yukawa couplings (see e.g.~\cite{Brax:1994ae}) and are 
typically generation-independent. The reason is that to obtain a 
non-trivial structure of the Yukawa couplings and/or CP violation at 
the renormalizable level often requires the quark fields of different 
generations to belong to the same twisted sector\footnote{This is always 
true in prime orbifolds since there is only one twisted sector.} 
(see e.g. \cite{Casas:1989qx,Lebedev:2001qg,Khalil:2001dr}).  
Consequently, these quark fields have the same modular weights.  
(In any case, the modular weights can only be $-1$ or $-2$ for 
non-oscillator states \cite{Ibanez:1992hc}. Oscillator states usually
correspond to SM singlets.) As a result, in these models
the K\"ahler metric is generation--independent.
The Yukawa couplings at $\langle T\rangle\sim1$ are given by
\begin{equation} 
{\cal Y}_{\alpha\beta\gamma}\sim e^{-\kappa_{_{\alpha\beta\gamma}}T}
\end{equation}
with order one coefficients $\kappa_{\alpha\beta\gamma}$,
and a nontrivial flavour structure of the $A-$terms results from
\begin{equation}
\Delta{\cal A}_{\alpha\beta\gamma}\sim\kappa_{{\alpha\beta\gamma}}~
{\cal Y}_{\alpha\beta\gamma} F^T\;.
\end{equation}
Analogous results hold for the Yukawa matrices generated by 
the Froggatt--Nielsen mechanism, in which case $U(1)$ charges play
the role of $\kappa_{\alpha\beta\gamma}$.

Similarly, in semirealistic brane models (for a review, see \cite{Blumenhagen:2005mu})
the K\"ahler metric for matter 
fields is often diagonal and generation--independent. For instance, 
replication of families naturally appears in intersecting brane models 
with different generations located at different intersections of the 
same branes \cite{Blumenhagen:2000wh}-\cite{Cremades:2003qj}. The K\"ahler 
metric is then diagonal and depends on the intersection angles (and 
moduli) \cite{Lust:2004cx} but is the same for fields of the same type 
belonging to different generations. Thus, the situation here is similar 
to the one in the heterotic string case.

The resulting soft masses are generation--independent at the string scale
(although they can generally be different for up and  down squarks, and 
for left and right squarks), while the $A-$terms can have a rich flavour 
structure due to Eq.~(\ref{eqn:A-non}). Thus, the string scale soft 
breaking lagrangian is parametrized by
\begin{equation}
(m^2_{Q,U,D})_{ij} = m^2_{Q,U,D}~\delta_{ij}~,\phantom{aaa}A_{ij}^{u,d} \;,
\label{eqn:univ_masses}
\end{equation}
in addition to flavour-independent parameters.

\subsection{Additional flavour violation in the K\"ahler potential only}

If the K\"ahler metric has a non-trivial generation dependence, e.g.
\begin{equation}
F^m \partial_m \tilde K_{\bar\alpha\beta}\not\propto 
\tilde  K_{\bar\alpha\beta}~,
\end{equation}
both the scalar masses and the 
$A-$terms have a non-trivial flavour structure. This situation can occur, 
for example, in the heterotic string if the quark field modular weights 
$n_\alpha$ are generation--dependent (see Eq.~(\ref{eqn:K})). 
The non--trivial flavour structures arise then from 
\begin{eqnarray}
&&\Delta{\mathpzc m}_{\bar\alpha\beta}^2 = n_{\bar\alpha} 
\tilde K_{\bar\alpha\beta} ~{\vert F^T\vert^2\over(T+\bar T)^2} 
\;,\nonumber\\
&&\Delta{\cal A}_{\alpha\beta\gamma} = -(n_\alpha+n_\beta+n_\gamma)
{\cal Y}_{\alpha\beta\gamma}~         
{F^T\over T+\bar T} ~{\hat W^\ast\over\vert\hat W\vert}e^{\hat K/2}  \;.
\label{eqn:m-A}
\end{eqnarray}
Here the modular weights are order one integers. Their typical values
are $-1$ and $-2$ due
to non--oscillator nature of the SM matter.

The K\"ahler metric is the only source of SUSY flavour violation if 
\begin{equation}
F^m \partial_m{\cal Y}_{\alpha\beta\gamma}\propto
{\cal Y}_{\alpha\beta\gamma}~.
\end{equation}
For example, the contribution 
$F^m \partial_m {\cal Y}_{\alpha\beta\gamma}$ 
to the $A-$terms vanishes if the Yukawa 
 structure is generated as in the 
Froggatt-Nielsen models \cite{Froggatt:1978nt} by non-renormalizable couplings through the 
vacuum expectation value of a scalar field $\phi$ which does 
not break supersymmetry. One then has
\begin{equation}
{\cal Y}_{\alpha\beta\gamma}={\cal Y}_{\alpha\beta\gamma}(\phi)~, 
\phantom{aaa}F^\phi=0 \;.
\label{eqn:FN}
\end{equation}

Since in the case discussed above the K\"ahler metric is diagonal, the 
soft mass terms are also diagonal, but generically nonuniversal. In 
contrast, the $A-$terms can be quite complicated due to the generation 
dependence of the modular weights (see e.g. \cite{Khalil:1999zn}). More 
complicated, non-diagonal soft mass terms can be obtained, for instance, 
in compactifications of the ten dimensional heterotic string  
on $(T_2/Z_3)^3$ orbifolds, where $T_2/Z_3$ are compact complex spaces
(``planes'') obtained by dividing complex tori $T_2$ by a discrete group $Z_3$.
In this case, the three generations of untwisted matter superfields 
can be associated with the  three ``planes'' $T_2/Z_3$ according to their
holomorphic indices. 
The K\"ahler metric is then non-diagonal and is given by
\begin{equation}
\tilde K_{\bar\alpha\beta}=({\rm Re}M_{\bar\alpha\beta})^{-1} \;,
\end{equation}
where $\alpha$, $\beta=1,2,3$ and the 9 moduli $M_{\bar\alpha\beta}$ 
parametrize the sizes of the compactification tori and the angles 
between the three ``planes'' $T_2/Z_3$ (see e.g. \cite{Casas:1989qx}). 
In such models, the soft terms will have a richer non-diagonal flavour 
structure depending on specific values of the moduli.

An interesting possibility is that non-trivial 
flavour dependence drops out of  the $A-$terms, but remains
in the soft mass terms. This occurs, for instance, if the vacuum
expectation values of the moduli fields take on special values  such that
\begin{eqnarray}  
F^m\partial_m \tilde K_{\bar\alpha\beta}\propto \tilde K_{\bar\alpha\beta} ~,
\phantom{aaa} 
F^m\partial_m {\cal Y}_{\alpha\beta\gamma}\propto 
{\cal Y}_{\alpha\beta\gamma}\nonumber \;,
\end{eqnarray}
\begin{eqnarray}  
F^m  \bar  F^{\bar n} \partial_m   \partial_{\bar n}\tilde K_{\bar\alpha\beta}\not\propto 
\tilde K_{\bar\alpha\beta}
\label{eqn:specialK}
\end{eqnarray}
and the flavour dependence appears only through the second derivatives. 
Then, the $A-$terms are proportional to the Yukawa couplings
while the soft masses are general.

We conclude that flavour violation through the K\"ahler potential 
generally leads to a complicated non-diagonal structure of the soft 
terms. The squark mass matrices and the $A-$terms are correlated, 
although this correlation can be far from transparent. The string 
scale soft breaking lagrangian is parametrized by 
\begin{equation}
(m^2_{Q,U,D})_{ij}~,\phantom{aa}A_{ij}^{u,d} \;.
\end{equation}
Two interesting special cases are: (i) diagonal soft masses with 
general $A-$terms, 
\begin{equation}
(m^2_{Q,U,D})_{ij} = (m^2_{Q,U,D})_i~\delta_{ij}~,
\phantom{aa}A_{ij}^{u,d} \;,
\end{equation}
as in  Eq.~(\ref{eqn:m-A}), and (ii)
general soft masses with universal $A-$terms, 
\begin{equation}
(m^2_{Q,U,D})_{ij}~,\phantom{aa}A_{ij}^{u,d}=A^{u,d}~Y_{ij}^{u,d} \;,
\label{zzz}
\end{equation}
as in Eq.~(\ref{eqn:specialK}).

\subsection{Both additional sources present}

This is a general case and not much can be said here apart from what 
already appears in Eq.~(\ref{eqn:mA}). Special cases have been   covered 
in previous subsections.
The string scale soft breaking lagrangian is general and 
is parametrized by  
\begin{equation}
(m^2_{Q,U,D})_{ij}~,\phantom{aa}A_{ij}^{u,d} \;.
\end{equation}

\subsection{Summary of the textures}

The above discussion leads us to the following supergravity benchmark 
textures:
\begin{eqnarray}
{\rm (A)}:~&&{\rm complete~universality:} \nonumber\\
           &&m^2_0,~A_0  \nonumber\\
{\rm (B)}:~&&{\rm generation~independent~
             scalar~masses~and~general}~A-
             {\rm terms:}\nonumber\\ 
&&m^2_Q,~m^2_U,~m^2_D,~ A_{ij}^u,~A_{ij}^d \nonumber\\
{\rm (C)}:~&&{\rm diagonal~scalar~masses~and~universal}~
              A-{\rm terms:}\nonumber\\ 
&&(m^2_Q)_i,~(m^2_U)_i,~(m^2_D)_i,~ A^u,~A^d \nonumber\\
{\rm (D)}:~&&{\rm diagonal~scalar~masses~and~general}~
              A-{\rm terms:}\nonumber\\ 
&&(m^2_Q)_i,~(m^2_U)_i,~(m^2_D)_i,~ A_{ij}^u,~A_{ij}^d \; \nonumber\\ 
{\rm (E)}:~&&{\rm general~soft~terms:}\nonumber\\
&&(m^2_Q)_{ij},~(m^2_U)_{ij},~(m^2_D)_{ij},~A_{ij}^u,~A_{ij}^d \;.
\nonumber
\end{eqnarray}
Textures (D) and (E) also include the possibility that the soft 
masses and the $A-$terms are correlated, as in the case of flavour
violation through the K\"ahler potential (Eq.~(\ref{eqn:m-A})).
As texture (C) we choose a restricted version of the Ansatz (\ref{zzz})
with diagonal squark masses.
Here we neglect supergravity radiative corrections which can be
considerable in certain models \cite{Choi:1997de}.

In what follows, we will study experimental constraints on the above 
textures and discuss how to distinguish them. They serve as boundary 
conditions at high energies and evolve with the energy scale. At the 
electroweak scale, each texture leads to a specific pattern of the mass 
insertions (or, more generally, flavour matrices at the interaction 
vertices). The main features of the resulting patterns  can be summarized
as follows: 
\vskip0.2cm
(A): very little flavour changing

(B): small LL, RR and significant LR, RL  flavour changing  

(C): substantial LL, RR and small LR, RL flavour changing

(D): substantial LL, RR, LR, RL flavour changing

(E): substantial LL, RR, LR, RL flavour changing
\vskip0.2cm
Here LL and RR refer to  chirality conserving flavor changing
transitions in the left-- and right--handed sectors, respectively.
LR and RL refer to chirality flipping flavor changing transitions.
For  textures (C) and (D), LL/RR flavour changing results from
non--universality of the soft masses at the string scale,
that is, their departure from the form 
(\ref{eqn:univ_masses}). Note that order one non--universality
applies to  $m^2_i$ rather than $m_i$, which makes a considerable
difference for the FCNC analysis.

Since different physical processes are sensitive to different types of 
mass insertions, the above textures (perhaps except for (D) and (E)) are 
distinguishable given enough experimental information.

\section{Low energy effects  of the textures}
\label{sec:textures}

In this section, we recall the steps which are necessary to obtain
the low energy manifestations of textures (A) to (E) and discuss their
main consequences. First of all, as we have already mentioned, each 
texture evolves with the energy scale and this evolution is described by 
the renormalization group (RG) equations. Different soft supersymmetry 
breaking terms evolve differently. The evolution from the GUT scale down 
to the electroweak scale mainly amounts to adding flavour-universal 
contributions to the squark mass matrices and the A-terms. These are due 
to gluino loops and grow with the gluino mass. 
The main effect of these contributions is that the average squark masse $\tilde M$
increases significantly. This has two important consequences: 
firstly, the mass insertions decrease as $1/\tilde M^2$ and, secondly,
 the bounds on the mass insertions relax as  $\tilde M$ or $\tilde M^2$.
Both of these effects make the FCNC problem milder \cite{Choudhury:1994pn}.

To deal with complicated flavour structures it is convenient to employ 
the mass insertion approximation \cite{Hall:1985dx} (although sometimes 
it may not be precise enough). The mass insertions are defined in the 
super-CKM basis, i.e. the basis in which the quark mass matrices are 
diagonal and positive,
\begin{eqnarray}
&&Y^u \longrightarrow V_L^{u\dagger}~ Y^u~  
V_R^{u}={\rm diag}(h_u,h_c,h_t)\;,\nonumber\\
&&Y^d \longrightarrow V_L^{d\dagger} ~Y^d~
V_R^{d}={\rm diag}(h_d,h_s,h_b)\;,
\end{eqnarray}
where $h_i$ denote the physical quark Yukawa couplings. To preserve the 
diagonal flavour structure of the supergauge vertices, the squark fields are 
rotated in the same fashion as the quark fields.
Thus, we have the following superfield 
transformations:
\begin{eqnarray}
&& \hat U_{L,R} \longrightarrow V^u_{L,R}~ \hat U_{L,R} \;, \nonumber\\
&& \hat D_{L,R} \longrightarrow V^d_{L,R}~ \hat D_{L,R} \;.
\end{eqnarray}
In this basis, the mass insertions at the electroweak scale are given by
\begin{equation}
(\delta^u_{XY})_{ij}\equiv
{\left({\cal M}^u_{XY}\right)^2_{ij}\over\tilde M^2}~,\phantom{aa}
(\delta^d_{XY})_{ij}\equiv
{\left({\cal M}^d_{XY}\right)^2_{ij}\over\tilde M^2}~,\phantom{aa}
i\neq j~,
\end{equation}
where ${\cal M}^{u,d}_{LL}$, ${\cal M}^{u,d}_{RR}$, 
${\cal M}^{u,d}_{LR}$ and ${\cal M}^{u,d}_{RL}$ are the $3\times3$ 
blocks of the full up and down squark mass squared matrices (see e.g. \cite{Misiak:1997ei})
and $\tilde M$ is the ``average'' squark mass appropriate for a given
mass insertion.
The mass insertion approximation 
works well when the mass insertions are significantly 
smaller than unity and the splittings among the eigenvalues of the mass 
matrix are significantly smaller than the eigenvalues themselves. 
The squark propagator in the mass insertion approximation has an 
expansion
\begin{eqnarray}
\langle\tilde q_\alpha\tilde q_\beta^\ast\rangle &=& 
{i\over\Bigl(k^2{\bf 1}-\tilde M^2{\bf 1}-\delta m^2\Bigr)_{\alpha\beta}}
\\
&=&{i\over k^2-\tilde M^2}~{\bf 1}_{\alpha\beta}~+~
{i\over(k^2-\tilde M^2)^2}~ \delta m^2_{\alpha\beta} ~+~  
{i\over(k^2-\tilde M^2)^3}~\delta m^2_{\alpha \gamma} ~
\delta m^2_{\gamma\beta}~+~\dots,\nonumber
\end{eqnarray}
where $\alpha,\beta$ are indices of the 6$\times$6 mass matrices. 
If the linear in $\delta m^2_{\alpha\beta}$ term happens to vanish for 
some ${\alpha,\beta}$, the leading contribution is provided by the 
``effective'' mass insertion $\tilde \delta_{\alpha\beta}\sim 
\delta_{\alpha\gamma}\delta_{\gamma\beta}$. However, this combination 
is not completely equivalent to a single mass insertion 
$\tilde\delta_{\alpha\beta} $ due to a different momentum
dependence of the relevant loop integral.

It is important to note that since the super-CKM basis is  defined 
only up to a phase, one must also fix the CKM phase convention 
\cite{Lebedev:2002wq}, which we take to be of the Wolfenstein type. 

Let us now discuss some features of SUSY flavour structures
in the super-CKM basis.

\subsection{$A-$terms in the super-CKM basis}

In order to study the  LR sector, it is convenient to factor out the 
Yukawa couplings from the $A-$terms (see Eq.~(\ref{eqn:mA})),
\begin{equation}
A^{u,d}_{ij}\equiv\tilde A^{u,d}_{ij} ~Y^{u,d}_{ij}\;.
\label{atilde}
\end{equation}
Deviations from universality are then encoded in the matrix
$\tilde A^{u,d}_{ij}$, which in the universal case has  all
entries equal: $\tilde A^{u,d}_{ij}=\tilde A^{u,d}$. 

Unlike $A^{u,d}_{ij}$, matrix elements of $\tilde A^{u,d}_{ij}$ 
are typically ${\cal O}(1)$ times an overall scale factor \cite{Abel:2001cv}, which is 
determined by the SUSY breaking scale and is usually of the order of 
the gravitino mass $m_{3/2}$.
For instance, in the heterotic string $Y_{ij}\sim e^{-\alpha_{ij}T}$
with $\alpha_{ij}$ of order one and $T$ being the vacuum expectation
value of the $T-$modulus. The non-universality of $\tilde A_{ij}$ is 
in this case given by $\partial_T\ln Y_{ij}\sim\alpha_{ij}$, or, for 
non-universal modular weights $n_i$, by $n_i+n_j$. In the case of the 
Froggatt-Nielsen mechanism, the role of $\alpha_{ij}$'s is played by 
the  $U(1)_X$ charges, with the same conclusion
\cite{Dudas:1995eq}-\cite{Ross:2002mr}. The amount of non-universality 
may reduce for democratic Yukawa textures, due to smaller values of the 
effective $\alpha_{ij}$'s, but this is a model-dependent issue.

The $A-$terms undergo RG evolution to low energies with the dominant 
contribution coming from gluino loops, which has an ``aligning'' effect 
similar to those for the squark mass matrices. Upon going over to the 
super-CKM basis, the $A-$terms transform just as  the Yukawa matrices do,
\begin{eqnarray}
&&A^u_{ij} \longrightarrow\left(V_L^{u\dagger}~A^u~V_R^{u}\right)_{ij}
\;,\nonumber\\
&&A^d_{ij} \longrightarrow\left(V_L^{d\dagger}~A^d~V_R^{d}\right)_{ij}\;.
\end{eqnarray}
In this basis, $A^{u,d}_{ij}$ are generally non-diagonal and 
the resulting flavour-changing mass insertions are ($i\not=j$):
\begin{equation}
\left(\delta_{LR}^{u,d}\right)_{ij}= {v_{u,d}~A_{ij}^{u,d}
\Bigl\vert_{SCKM}\over\tilde M^2} \;,
\end{equation}
where $v_{u,d}$ are the Higgs VEVs and $\tilde M$ is the appropriate
average squark mass at the electroweak scale. To discuss orders of 
magnitude of various insertions it is instructive to write
\begin{equation}
A^u_{ij}\Bigl\vert_{SCKM}=
{\rm scale ~factor}\times\Bigl(\alpha_{ij}~h_u +
\beta_{ij}~h_c +\gamma_{ij}~ h_t\Bigr) \;, 
\label{eqn:Asckm}
\end{equation}
and similarly for the down sector. Here $\alpha_{ij}$, $\beta_{ij}$,
$\gamma_{ij}$ are model dependent $\leq{\cal O}(1)$ coefficients
parametrizing departure from universality. 
In the universal case,
\begin{eqnarray}
&& \alpha_{11}=1 {\rm ~~and~~0~otherwise} \;,\nonumber\\
&& \beta_{22}=1 {\rm ~~and~~0~otherwise} \;,\label{eqn:abg}\\
&& \gamma_{33}=1 {\rm ~~and~~0~otherwise} \;.\nonumber
\end{eqnarray}
This representation of the $A-$terms is useful for estimating typical 
magnitudes of the LR mass insertions. Since $\tilde A_{ij}^{u,d}$ are 
of order unity, $A_{ij}^{u,d}$ have a similar structure to that of the 
Yukawa matrices and the misalignment is characterized by deviation of 
$\alpha_{ij}$, $\beta_{ij}$, $\gamma_{ij}$ from the universal limit 
(\ref{eqn:abg}). This deviation is expected to be small if the 
Yukawa matrices and the $A-$terms are diagonalized by small angle
rotations, which is often the case for hierarchical Yukawa textures.
This ceases to be true in the case of democratic textures.

An interesting limiting case is  {\it matrix-factorizable} $A-$terms, 
i.e. such that they can be written as
\begin{equation} 
A^{u,d}\equiv\tilde B^{u,d}\cdot Y^{u,d}+Y^{u,d}\cdot\tilde C^{u,d}
\end{equation}
in the matrix sense with $\tilde B^{u,d}_{ij}$ and 
$\tilde C^{u,d}_{ij}\simlt{\cal O}(1)$ times a scale factor.
In this case,
\begin{equation}
A^u\Bigl\vert_{SCKM}=
\tilde B^u\Bigl\vert_{SCKM}\cdot ~{\rm diag}(h_u,h_c,h_t) +
{\rm diag}(h_u,h_c,h_t)\cdot\tilde C^u\Bigl\vert_{SCKM}\;
\end{equation}
and similarly for $A^d$. 
Here $\tilde B^u\Bigl\vert_{SCKM}$ and $\tilde C^u\Bigl\vert_{SCKM}$ 
have elements of order $\simlt{\cal O}(1)$, again up to an overall 
scale. This implies, for instance, that the (12) element contains 
contributions only from $h_u$ and $h_c$, the (11) element -- only from 
$h_u$, etc., and
\begin{equation}
A^u_{ij}\Bigl\vert_{SCKM}
\leq{\rm scale ~factor}\times({\cal O}(1)~ h_i +{\cal O}(1)~ h_j)\;.
\label{eqn:eq1}
\end{equation}
This limits the magnitude of the LR mass insertions for the first two 
generations and makes the SUSY FCNC problem less severe. Obviously, 
this situation occurs in the universal case. Other examples include 
models in which the Yukawa hierarchy is produced via a Froggatt-Nielsen 
field  and models with non--universal modular weights (Eq.\ref{eqn:m-A}),
such that $\tilde A_{ij}=a_i+b_j$ (see also \cite{Kobayashi:2000br}). 
We also note that the form (\ref{eqn:eq1}) is favoured by the absence 
of charge and color breaking minima in the scalar potential 
\cite{Casas:1996de}.

Finally, a useful estimate of the mass insertions is obtained by setting 
the overall scale of the $A-$terms (and the $\mu$--term) to be equal to 
the average squark mass. Then, for the up sector,
\begin{equation}
(\delta_{LR}^u)_{ij}\sim\alpha_{ij} {m_u\over\tilde M} +
\beta_{ij}{m_c\over\tilde M} + \gamma_{ij}{m_t\over\tilde M} \;.
\label{eqn:LRest}
\end{equation}

\subsection{LL and RR sectors in the super--CKM basis}

Upon going to the super-CKM basis, the LL and RR blocks
of the squark mass matrix are rotated as
\begin{eqnarray}
({\cal M}^{u,d})^2_{LL} \longrightarrow 
V_L^{u,d~\dagger}({\cal M}^{u,d})^2_{LL}~V_L^{u,d}\;,\nonumber\\
({\cal M}^{u,d})^2_{RR} \longrightarrow 
V_R^{u,d~\dagger}({\cal M}^{u,d})^2_{RR}~V_R^{u,d}\;.
\end{eqnarray}
The squark mass squared matrices $({\cal M}^{u,d})^2_{LL}$ and 
$({\cal M}^{u,d})^2_{RR}$ at the electroweak scale are determined by 
the original textures (A)-(E) and by the RG evolution. For  
textures (A)-(D), these matrices are diagonal and therefore remain 
approximately  diagonal after the RG evolution. Flavour violation at 
the electroweak scale is due to the rotations to the super-CKM
basis. We can easily estimate the order of magnitude of the expected
effects. Let us assume for a moment that the first two and the third
generations do not communicate and consider the (12) block,
\begin{equation}
{\cal M}^2_{LL}=\left(\matrix{m^2_1 & 0\cr0 & m^2_2}\right) \;.
\label{eqn:mLL12}
\end{equation}
Parametrizing the orthogonal rotation matrix $V_L$ by $\cos\theta$
and $\sin\theta$ in the usual fashion, we get
\begin{equation}
(\delta_{LL})_{12}= \cos\theta ~\sin\theta~ {\Delta m^2\over\tilde M^2}
\label{deg}
\end{equation}
in the super-CKM basis, with $\Delta m^2_{}\equiv m^2_1-m^2_2$ and 
$\tilde M^2\equiv{1\over2}(m^2_1+m^2_2)$. Thus, small mass insertions 
are obtained  for nearly degenerate masses $m_1$ and $m_2$ and/or 
for a small rotation angle $\theta$.

Consider now the case of 3 generations with the first two being 
degenerate,
\begin{equation}
{\cal M}^2_{LL} = \left(\matrix{m^2_1 &   0    & 0 \cr
                                0     &  m^2_1 & 0 \cr
                                0     &   0    & m^2_3}\right) \;.
\end{equation}
When the rotation matrix $V_L$ is well approximated by the CKM matrix, 
we have
\begin{eqnarray}
&&(\delta_{LL})_{12}\sim10^{-4}~e^{i{\cal O}(1)}~
{\Delta m^2_{}\over\tilde M^2} \;,\nonumber\\
&&(\delta_{LL})_{13}\sim10^{-3}~e^{i{\cal O}(1)}~
{\Delta m^2_{}\over\tilde M^2} \;,\\
&&(\delta_{LL})_{23}\sim 10^{-2}~e^{i{\cal O}(10^{-1})}~
{\Delta m^2_{}\over\tilde M^2} \;,\nonumber
\end{eqnarray}
where $\Delta m^2_{}\equiv m^2_1-m^2_3$ and $\tilde M$ is the average squark
mass.
This gives a good idea of the  expected magnitude of mass 
insertions in the case of a small angle rotation, but can be drastically
different for textures requiring a large angle rotation.

As we will see, FCNC constraints require mass insertions in the LL and 
RR sectors to be quite small but, nevertheless, leave  room for 
departures from degeneracy of the eigenvalues, particularly for small 
rotation angles.

In the most general texture (E), the flavour off-diagonal entries of the 
squark mass squared matrices are present already at the high energy 
scale. Their RG evolution is not important, so at the electroweak scale 
they remain  of similar order of magnitude. 
Furthermore, barring accidental cancellations, the rotations to the 
super-CKM basis do not change the qualitative picture. 
The only important effect is the increase of the flavour--diagonal entries
due to the RG running,
which reduces the magnitude of the mass insertions.
Taking this  into account, experimental
constraints on the off--diagonal LL and RR mass insertions can be 
applied directly to the high--energy texture (E). The result is that
such insertions have to be small and generic textures (E)
are inconsistent with experiment. 

It is  possible that the soft breaking terms and the Yukawa matrices
align due to some horizontal symmetry, resulting in suppressed mass 
insertions. In this paper, we will take a conservative view and will 
not pursue this option further. 

\subsection{Yukawa textures}

An important issue to address is dependence of SUSY FCNC on Yukawa 
textures. To cover both ends of the spectrum, we take a few 
representative examples with both small and large angle rotations.
\vskip0.3cm

{\bf (i). The simplest texture.}\\
The simplest texture contains no extra parameters beyond those
already present in the CKM matrix and quark masses:\footnote{An alternative
texture of this sort, $Y^u\propto V_{CKM}^\dagger{\rm diag}~(m_u,m_c,m_t)$, 
$Y^d\propto{\rm diag}(m_d,m_s,m_b)$ would lead to  smaller FCNC
effects in the down type quark sector, which is constrained by experiment
stronger than the up sector.}
\begin{eqnarray}
&&Y^u = {1\over v \sin\beta} ~{\rm diag} (m_u,m_c,m_t) \;,\nonumber\\
&&Y^d = {1\over v \cos\beta} ~V_{CKM}^\dagger~{\rm diag} (m_d,m_s,m_b) 
\;,
\end{eqnarray}
with $v^2=v_u^2+v_d^2$ and $\tan\beta=v_u/v_d $. To diagonalize these 
matrices, only a left handed rotation $V_{CKM}$ in the down sector is 
required.
\vskip0.3cm

{\bf (ii). A hierarchical texture.}\\ 
Hierarchical textures encode the hierarchy of the quark masses in a 
transparent manner and arise in various models. For example, they appear 
in string models due 
to the exponential dependence of the twisted sector couplings on moduli 
(for a recent analysis, see \cite{Ko:2004ic}) 
or due to a Froggatt-Nielsen type mechanism 
\cite{Faraggi:1993su}. 

For definiteness, we use a set of textures from 
Ref.~\cite{Chankowski:2005qp}. This is an $SU(5)\times U(1)$ model with  
order ${\cal O}(1)$ coefficients chosen so that a good fit to the 
fermion masses and mixings is assured. The $U(1)_X$ charges are 
$q=\bar u=\bar e=(3,2,0)$ and $\bar d= l=(4,2,2)$. The up-quark Yukawa 
has a structure
\begin{eqnarray}
&&Y^u=\left(\matrix{\epsilon^6 & \epsilon^5 & \epsilon^3 \cr
                           \epsilon^5 & \epsilon^4 & \epsilon^2 \cr
                           \epsilon^3 & \epsilon^2 & \epsilon^0}\right).
\end{eqnarray}
The down quark Yukawa matrix $Y^d$  is given by
\begin{eqnarray}
&&Y^d_{ij} = \left(C^{(1)} _{ij}+ \kappa\, C^{(2)}_{ij}\right) 
\epsilon^{q_i+\bar d_j} ~, 
\label{eqn:YdYe}
\end{eqnarray}
where $\kappa=0.3$, $\epsilon=0.22$ and matrices $C^{(1,2)}$ have a 
structure of the form
\begin{eqnarray}
&&C_{1,2}=\left(\matrix{\epsilon^7 & \epsilon^5 & \epsilon^5 \cr
                        \epsilon^6 & \epsilon^4 & \epsilon^4 \cr
                        \epsilon^4 & \epsilon^2 & \epsilon^2}\right)~.
\end{eqnarray}

Diagonalization of the Yukawa matrices in this model requires small 
angle rotations in the left-handed sector and large angle rotations
in the right-handed sector. The latter are correlated with the
large neutrino mixing angles in GUT models \cite{Chang:2002mq}.
\vskip0.2cm

{\bf (iii). A democratic texture.}\\
A strictly democratic texture predicts one massive and two massless 
quarks in the up and down sectors. 
Realistic quark masses and mixings can be produced by a
perturbation around  this texture
 \cite{Branco:1990fj,Fritzsch:1989qm}. Democratic textures can also arise in string 
models, e.g. when the exponential suppression of the twisted sector 
couplings is not significant (see e.g. Abel {\it et al.} in 
\cite{Cremades:2003qj}).

In our numerical analysis, we use democratic textures 
similar to those of Ref.~\cite{Abel:2002zg},
\begin{eqnarray}
Y^u={m_t\over 3 v \sin\beta} \left(\matrix{1.013&0.987&0.999\cr
                                           0.987&1.013&0.999\cr
                                           0.999&0.999&0.998}\right)\;,
\nonumber
\end{eqnarray}
\begin{eqnarray}
Y^d={m_b\over3v\cos\beta} ~K^\dagger\cdot
               \left(\matrix{0.987&0.905&0.968\cr
                             0.903&1.212&1.008\cr
                             0.967&1.008&1}\right)\cdot K \;, 
\end{eqnarray}
\begin{eqnarray}
K={\rm diag(1,e^{-0.01i},e^{0.01i} )} \;.\nonumber
\end{eqnarray}
These matrices are diagonalized by a large angle rotation.

\section{Experimental constraints}
\label{sec:expconstr}

Before we present our numerical analysis, in this section we collect 
experimental constraints on various mass insertions and also 
discuss qualitatively their implications for textures (A)-(E).
\vskip0.5cm

\noindent{\it Electric dipole moments}

\noindent The electric dipole moments of the neutron and mercury atom (and 
of the electron, in the lepton sector) are especially sensitive to 
flavour conserving LR mass insertions for the first generation. In 
particular, the current bounds
\begin{eqnarray}
&&\vert d_n\vert<6\times10^{-26} ~e~{\rm cm} \;, \nonumber\\
&&\vert d_{\rm Hg}\vert < 2\times10^{-28} ~e~{\rm cm} 
\end{eqnarray} 
impose the constraints \cite{Gabbiani:1996hi}-\cite{Abel:2001vy}:
\begin{equation}
\vert{\rm Im}(\delta_{11}^{\{ u,d \}})_{LR}\vert
\leq10^{-7}\div10^{-6} \;.
\label{edmbound}
\end{equation}
This implies that the corresponding CP phases of the $\mu-$ and $A-$terms
have to be small. Clearly, this SUSY CP problem \cite{Ellis:1982tk}
arises for all supergravity textures, including the flavour universal 
one. The unwanted CP phases can be suppressed in some special cases,
e.g. in the dilaton dominated SUSY breaking scenario with an axionic symmetry 
\cite{Lebedev:2002zt} or with the help of the CP phase alignment 
\cite{Ibanez:2004iv}, but this requires additional assumptions.

In the non-universal case, the problem becomes more severe due to 
additional CP phases which appear in the process of diagonalizing the 
quark mass matrices and the enhanced magnitude of the mass insertions 
\cite{Khalil:2001dr,Abel:2001cv}. For example, 
\begin{equation}
(\delta_{LR}^u)_{11}\sim
{\tilde A~(m_u+\varepsilon m_c+\varepsilon^\prime m_t)\over\tilde M^2}\;,
\end{equation}
where $\tilde A\sim\tilde M$ is the ``overall'' scale of the $A-$terms 
and $\varepsilon$, $\varepsilon^\prime$ are model-dependent coefficients.
This makes it more difficult to satisfy the bound (\ref{edmbound}). 
In this sense, the EDMs prefer some sort of universality, at least 
in the $A-$terms, although they are problematic in any case.

The problem could be partly solved by decoupling the first two sfermion 
generations (although the third generation still contributes to the EDMs 
through the Weinberg operator). However, it appears difficult to realize this 
possibility in specific models. Since the overall scale of the soft terms in 
SUGRA is given by $m_{3/2}$, a very large splitting between the masses of the 
first two and the third sfermion generations would require some sort of a 
singularity in the derivatives of the K\"ahler potential. In practice, this 
does not happen and the non-universality is governed, for instance, by order 
one modular weights (Eq.~(\ref{eqn:K})). Another option would be to assume a 
large $m_{3/2}$ and obtain  light, as required by naturallness, third 
generation sfermions via RG evolution to infrared fixed points. However, in 
this case it would be problematic to get light enough gauginos.

Given these difficulties, we will not attempt to resolve the SUSY CP 
problem and, in what follows,  will simply treat the EDMs as a 
constraint on all of the SUGRA textures.
\vskip0.5cm

\noindent{\it Kaon observables}

\noindent The most important observables in the Kaon system are the 
Kaon mass splitting, $\Delta M_K$, and the parameters $\epsilon$ and 
$\epsilon^\prime$ measuring CP violation in Kaon decays,
\begin{eqnarray}
&&\Delta M_K=M_{K_L}-M_{K_S}\simeq3.5\times10^{-15}{\rm ~GeV}\;,
\nonumber\\
&&\epsilon={A(K_L\rightarrow\pi\pi) \over A(K_S \rightarrow\pi\pi)}
\simeq 2.3\times10^{-3} \; ,
\nonumber\\
&&\epsilon^\prime/\epsilon=-{\omega\over\sqrt2\vert\epsilon\vert
{\rm Re}A_0}\Bigl({\rm Im}A_0 -{1\over\omega}{\rm Im}A_2\Bigr) 
\simeq1.9\times10^{-3}\;, \
\end{eqnarray}
where $A_{0,2}$ are the $K\rightarrow\pi\pi$ amplitudes for the 
$\Delta I=1/2,3/2$ transitions and 
$\omega\equiv{\rm Re }A_2/{\rm Re}A_0\simeq1/22$.

These observables place severe constraints on new physics flavour 
structures. In particular, SUSY contributions mediated by gluinos and 
squarks generally lead to $|\Delta M_K|$, $|\epsilon|$ and 
$|\epsilon^\prime/\epsilon|$ 
orders of magnitude too large. For gluino and squark masses of 500~GeV,
and assuming no accidental cancellations between contributions of 
different mass insertions, the measured $\Delta M_K$ imposes the 
following bounds on the mass insertions \cite{Gabbiani:1996hi},
\begin{eqnarray}
&&\sqrt{\Bigl\vert{\rm Re}\left(\delta_{12}^d\right)_{LL}^2\Bigr\vert}
< 4\times 10^{-2} \;, \phantom{aaa}
\sqrt{\Bigl\vert{\rm Re}\left(\delta_{12}^d\right)_{LL}
\left(\delta_{12}^d\right)_{RR}\Bigr\vert}
< 3\times10^{-3} \;, \nonumber\\
&&\sqrt{\Bigl\vert{\rm Re}\left(\delta_{12}^d\right)_{LR}^2 
\Bigr\vert}< 4\times 10^{-3} \;,
\end{eqnarray}
while the measured value of the $\epsilon$ parameter \cite{Gabbiani:1996hi}
imposes the bounds 
\begin{eqnarray}
&&\sqrt{\Bigl\vert{\rm Im}\left(\delta_{12}^d\right)_{LL}^2\Bigr\vert}
<3\times10^{-3} \;, \phantom{aaa}
\sqrt{\Bigl\vert{\rm Im}\left(\delta_{12}^d\right)_{LL}
\left(\delta_{12}^d\right)_{RR}\Bigr\vert}
<2\times10^{-4} \;, \nonumber\\
&&\sqrt{\Bigl\vert{\rm Im}\left(\delta_{12}^d\right)_{LR}^2\Bigr\vert}
<4\times 10^{-4} \;.
\end{eqnarray}
The bounds on the RR mass insertion are the same as the ones 
on the LL insertion. All these bounds scale as $\tilde M$ when 
the squark masses are changed. Furthermore,
the chargino-squark loop contributions to $\Delta M_K$ and $\epsilon$
impose analogous (although somewhat weaker) constraints on the 
up squark sector mass insertions \cite{Khalil:2001wr}. For $\mu=M_2=350$ GeV 
and the squark masses of 500 GeV, they read
\begin{eqnarray}
\sqrt{\Bigl\vert{\rm Re}\left(\delta_{12}^u\right)_{LL}^2\Bigr\vert}
<1\times 10^{-1}~,\phantom{aaa}
\sqrt{\Bigl\vert{\rm Im}\left(\delta_{12}^u \right)_{LL}^2\Bigr\vert}
< 1\times 10^{-2}\;.
\end{eqnarray}
Similar constraints involving RR and LR mass insertions are much weaker 
because of the factors $m_q/M_Z$ suppressing the couplings of the
right-chiral squarks to charginos and quarks.
Finally, for the same gluino and down type squark masses as above, the 
measured value of the $\epsilon^\prime$ parameter sets the rough bounds 
\cite{Gabbiani:1996hi}
\begin{eqnarray}
&& \Bigl\vert{\rm Im}\left(\delta_{12}^d \right)_{LL}\Bigr\vert
<5\times10^{-1} ~,\phantom{aaa}
\Bigl\vert{\rm Im}\left(\delta_{12}^d\right)_{LR}\Bigr\vert<2\times10^{-5} \;.
\end{eqnarray}
{}In addition, from the chargino--up type squark contributions to 
$\epsilon^\prime$ one obtains a rather weak limit
$\Bigl\vert{\rm Im}\left(\delta_{12}^u\right)_{LL}\Bigr\vert<0.3$
\cite{Khalil:2001wr} and essentially no bound on  other mass  
insertions.

Let us now discuss implications of these constraints. Clearly, at the 
electroweak scale only little  
mixing between squarks of the first two generations is allowed.
The strongest bounds on the chirality conserving mass insertions 
come from $\Delta M_K$ and $\epsilon$, while those on the chirality changing  
mass insertions are due to $\epsilon^\prime$.

In the LL and RR sectors, the allowed mass insertions are of order $10^{-2}$
or smaller. This means that the soft masses in the original basis are almost 
diagonal (barring alignment) and the diagonal entries are almost degenerate
at low energies. In the down sector, the most conservative bounds imply that 
this degeneracy is at a percent level for democratic Yukawa textures and at 
about 10\% level for hierarchical textures. In the up sector, the mass 
splittings can be larger\footnote{If the CKM matrix is entirely due to the 
up sector and the Yukawa textures are hierarchical, the constraints are 
particularly weak and no significant degeneracy is required. This, however, 
appears to be a rather special case.}.
This implies that the LL and RR sectors are to some extent universal at
low energies. As we will see in the next section, the RG running from the GUT 
scale to the electroweak scale has an important ``aligning'' effect. As a  
result, constraints on the high energy values of the soft parameters are 
milder.

The situation is very different in  the LR sector. Consider the mixing of 
the first two generation squarks  and assume that their mixing 
with the third generation squarks is small, $\leq$$10^{-2}$. A natural 
magnitude of the mass insertion is then (see Eq.~(\ref{eqn:LRest})) 
\begin{equation} 
\left(\delta_{12}^d\right)_{LR}\sim\alpha_{12}{m_d\over \tilde M} +
\beta_{12} {m_s\over \tilde M}\simlt{\cal O}(10^{-4})\;, 
\label{eqn:K-K}
\end{equation}
where $\alpha_{12}$, $\beta_{12}$ are $\simlt{\cal O}(1)$ model dependent 
coefficients. It follows that all the bounds except for that from $\epsilon^\prime$ are 
satisfied automatically. $\epsilon^\prime$ imposes a rather mild constraint on 
the imaginary part of $A_{12}^d$ (see, e.g. \cite{Masiero:1999ub}). The same 
considerations also apply to the up squark sector. This means that order one 
non--universality is allowed in the $(12)$ block of $\tilde A^d_{ij}$.

The above estimate holds in a wide class of models including those with 
hierarchical textures, matrix--factorizable A--terms, etc. However, it may not 
apply to the case of democratic textures, which we study below numerically. 

Finally, we note that there also exist bounds on products of the mass 
insertions when one goes beyond a single mass insertion approximation. 
To give an example, chargino--squark penguin diagrams with two mass insertions 
on the squark line modify, in particular, the effective $Z^0\bar ds$ vertex. 
The resulting SUSY contribution to BR($K^+\rightarrow\pi^+\nu\bar\nu$) 
does not exceed its measured value provided 
$\vert{\rm Re}(\delta_{LR}^u)_{32}^\ast(\delta_{LR}^u)_{31}\vert<0.2$   
\cite{Colangelo:1998pm}. In SUGRA models, this product is expected to be
bounded by $(m_t/\tilde M)^2\sim10^{-1}$ and the constraint is satisfied 
automatically.
However, in other scenarios, 
large SUSY effects in the $Z^0\bar ds$ vertex are possible. They 
can lead, in particular, to the branching ratio of 
the $K_L\rightarrow\pi^0\nu\bar\nu$ decay  by up to two orders of magnitude larger 
than the SM prediction \cite{Colangelo:1998pm}.
\vskip0.5cm

\noindent {\it $D^0$-$\bar D^0$ mixing}

\noindent The experimental bound on the $D^0$-$\bar D^0$ mixing is
\begin{equation}
\Delta M_D < 4.8 \times 10^{-14}~ {\rm GeV} \;.
\end{equation}
For the gluino and squark masses of 500 GeV, the constraints on the up type 
squark mass insertions read \cite{Chang:2001ah}:
\begin{eqnarray}
&&\sqrt{\Bigl\vert{\rm Re}\left(\delta_{12}^u\right)_{LL}^2\Bigr\vert}
<5\times10^{-2} \;, \phantom{aaa}
\sqrt{\Bigl\vert{\rm Re}\left(\delta_{12}^u\right)_{LL}
\left(\delta_{12}^u\right)_{RR}\Bigr\vert}
<1\times10^{-2} \;, \nonumber\\
&&\sqrt{\Bigl\vert{\rm Re}\left(\delta_{12}^u\right)_{LR}^2\Bigr\vert}
<2\times10^{-2} \;,
\end{eqnarray}
and they scale as $\tilde M$ when the squark masses are changed. 
The above bounds result from the gluino-up type squark contribution to 
$\Delta M_D$ and are comparable to the ones stemming from
the chargino-squark contributions to the Kaon observables discussed in the 
previous section. 
Concerning the LR mass insertions in the up sector,
an estimate analogous to Eq.~(\ref{eqn:K-K}) holds 
and we reach the same conclusion that
order one non-universality in the (12) block of the $A-$terms is allowed. 
\vskip0.5cm

\noindent{\it $B$ meson observables}

The most important constraints on the flavour changing
transitions involving the $b$ quark are
\begin{eqnarray}
&&\Delta M_{B_d} \simeq 3.4\times10^{-13} ~{\rm GeV} \;,\nonumber\\
&&{\rm BR}(\bar B\rightarrow X_s\gamma) \simeq 3.3\times 10^{-4} \;, \\
&&S_{B^0_d\rightarrow\psi K_S} \simeq 0.73 \;,\nonumber
\end{eqnarray}
where $S_{B^0_d\rightarrow\psi K_S}$ measures the CP violating asymmetry in 
the $B^0_d \rightarrow\psi K_S$ decay, proportional to 
$\Gamma(\bar B^0_d\rightarrow\psi K_S)-\Gamma(B^0_d\rightarrow\psi K_S)$.

For the gluino and squark masses of 500 GeV, gluino--squark loop contributions
to $\Delta M_{B_d}$ lead to the following constraints on the 
down type squark  mass insertions \cite{Gabbiani:1996hi}:
\begin{eqnarray}
&&\sqrt{\Bigl\vert{\rm Re}\left(\delta_{13}^d\right)_{LL}^2\Bigr\vert}
<1\times10^{-1} \;, \phantom{aaa}
\sqrt{\Bigl\vert{\rm Re}\left(\delta_{13}^d\right)_{LL}
\left(\delta_{13}^d\right)_{RR}\Bigr\vert}
<2\times10^{-2} \;, \nonumber\\
&&\sqrt{\Bigl\vert{\rm Re}\left(\delta_{13}^d\right)_{LR}^2\Bigr\vert}
<3\times10^{-2} \;.
\end{eqnarray}
These limits scale as $\tilde M$ when the squark masses are changed. 
$S_{B^0_d\rightarrow \psi K_S}$ imposes  similar constraints 
on the imaginary parts of the same combinations of the mass 
insertions \cite{Gabrielli:2002fr}. 
As the value of $\Delta M_{B_s}$
is still not bounded from above by experiment, there are
no similar limits on the $23$ mass insertions. 
The experimental value of ${\rm BR}(\bar B\rightarrow X_s\gamma)$ sets 
a limit on the 
absolute value of the $(\delta^d_{23})_{LR}$ mass insertion \cite{Gabbiani:1996hi}:
\begin{equation}
\bigl\vert(\delta^d_{23})_{LR}\bigr\vert < 1.5\times10^{-2}
\label{eqn:bsg}
\end{equation}
for the gluino and squark masses of 500 GeV and the bound  scales as $\tilde M^2$ 
when the squark masses are changed. 

Constraints on the up sector mass insertions are quite weak: the 
chargino-squark contribution to $\Delta M_{B_d}$ leads  to the bound 
on $\left(\delta_{13}^u\right)_{LL}$ 
\cite{Gabrielli:2002fr}:
$\vert\left(\delta_{13}^u\right)_{LL}\vert\leq {\cal O} (10^{-1})$.
The insertions $\left(\delta_{13}^u\right)_{RR}$, 
$\left(\delta_{13}^u\right)_{LR}$, $\left(\delta_{13}^u\right)_{RL}$ as well
as all $\left(\delta_{23}^u\right)_{XY}$  are essentially 
unconstrained.

In SUGRA models, the LR mass insertions connecting the third 
generation with the other  two are generically of the order
\begin{eqnarray}
&&\left(\delta_{i3}^d\right)_{LR}\simlt{\cal O}\left({m_b\over 
\tilde M}\right)\sim10^{-2}\;, \nonumber\\
&&\left(\delta_{i3}^u\right)_{LR}\simlt{\cal O}\left({m_t\over 
\tilde M}\right)\sim0.1\div1\;
\label{deltai3}
\end{eqnarray}
for $i=1,2$ and the experimental constraints are satisfied automatically. 
Thus, the current bounds allow for order one non-universality in the $A-$terms
involving the third generation.
Some degree of universality is required
in the chirality conserving sectors, although the constraints 
are much weaker than those  on the mixing of the 
first two generations.

To conclude this section, the above simple estimates show that order one
nonuniversality in the $A-$terms is consistent with the current data, 
whereas the soft mass terms are required to be essentially
diagonal and somewhat degenerate at low energies. 
As explained earlier, the corresponding 
constraints on the high energy parameters are significantly
weaker due to the aligning effect of the RG evolution.
Below we 
confirm these conclusions numerically.

\section{Numerical results}

In this section, we present results of our numerical analysis for textures
(A), (B) and (C) and compare them with the experimental constraints listed 
in section \ref{sec:expconstr}. Barring accidental cancellations, these 
results also apply to texture (D) which is a combination of textures (B) 
and (C). For texture (E), no numerical analysis is needed to see that it 
is inconsistent with the FCNC constraints unless it reduces to one of the 
special cases (A)-(D). 

The numerical analysis of this section is necessary to support our qualitative 
conclusions of section 4. Moreover, it illustrates the dependence of the FCNC 
constraints on the chosen Yukawa textures and on squark and gluino masses.
It is also important to study at a quantitative level the still remaining
room for flavour dependence in the K\"ahler potential and  the $A-$terms
in supergravity models, so that prospects for further experimental 
investigations can be assessed.

In our analysis, we use the  Yukawa textures described in section 
\ref{sec:textures} and the standard 1-loop RG equations for the evolution 
of the soft terms from the high energy scale down to the electroweak scale.
Our results are presented as a function of the high (string) scale values 
of the parameters and compared with the limits on mass insertions. The limits
shown in the plots are properly rescaled to account for the actual values of 
low energy mass parameters obtained from the RG evolution.
\vskip0.3cm

\noindent{\it Texture (A)}

Flavour violating effects are very small, especially in the RR sector. For 
completeness, in Table \ref{table77} we provide representative mass 
insertions generated by the RG running. 
\begin{table}[h]
\begin{center}
\begin{tabular}{|c|c|c|c|}
\hline
mass insertion  & 12 & 13  & 23 \\
\hline
d, LL &  $10^{-4}$  &  $10^{-3}$  &  $10^{-2}$ \\
d, RR &  $10^{-13}$ &  $10^{-10}$ &  $10^{-8}$ \\
d, LR &  $10^{-8}$  &  $10^{-5}$  &  $10^{-5}$ \\
u, LL &  $10^{-7}$  &  $10^{-5}$  &  $10^{-4}$ \\
u, RR &  $10^{-16}$ &  $10^{-12}$ &  $10^{-9}$ \\
u, LR &  $10^{-11}$ &  $10^{-8}$  &  $10^{-6}$ \\
\hline
\end{tabular}
\end{center}
\caption{
RG generated mass insertions for $\tan\beta=3$, $m_0=200$~GeV,
$A_0=50$~GeV and $M_{1/2}=100$ GeV.
\label{table77}}
\end{table}

\noindent{\it Texture (B)}

\begin{figure}
\epsfig{figure=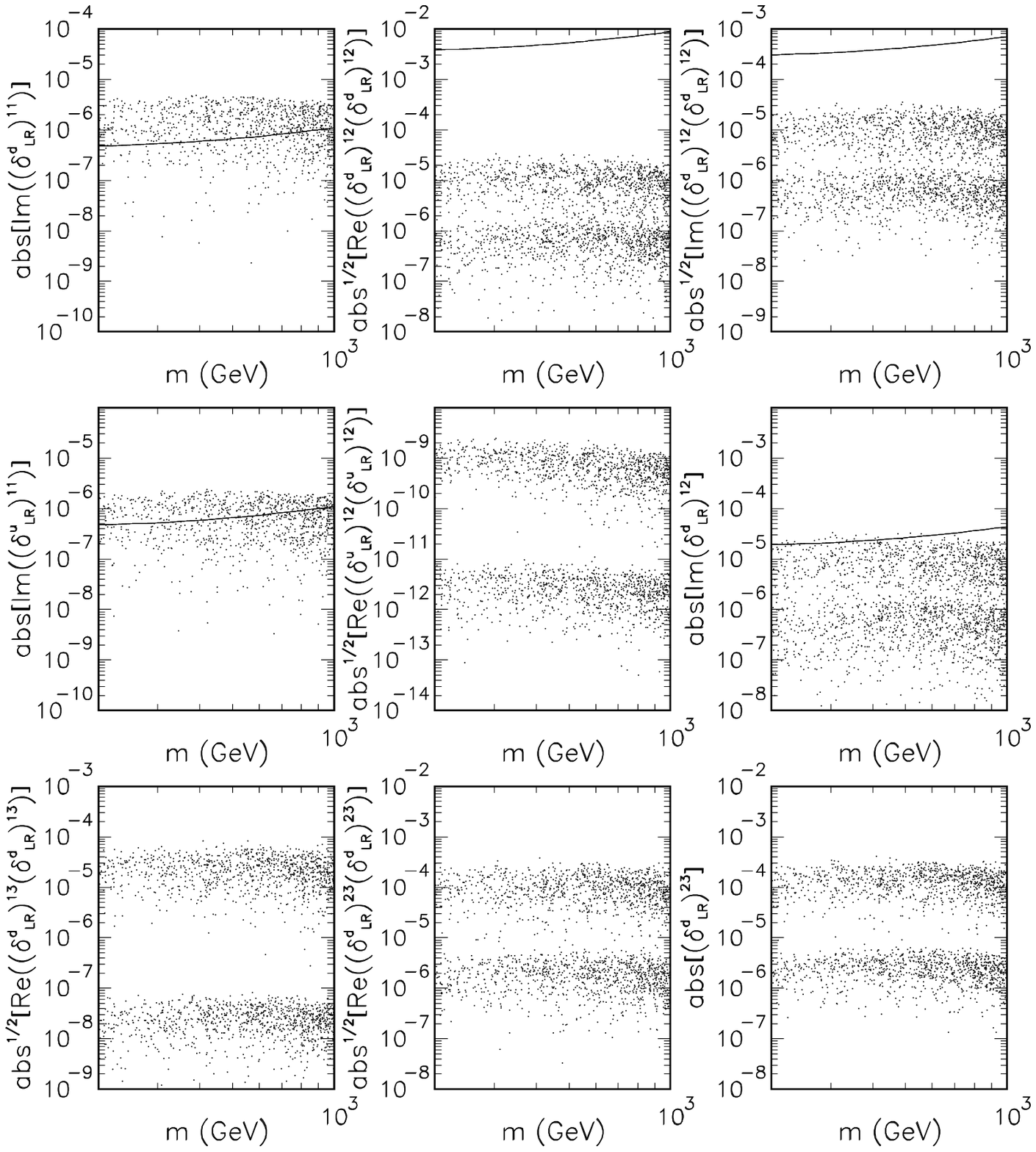,width=\linewidth} 
\vspace{1.0truecm}
\caption{
Phenomenologically relevant combinations of the insertions 
$(\delta^{d,u}_{LR})^{JI}$ and $(\delta^{d,u}_{RL})^{JI}$
for the hierarchical Yukawa texture (i) as a function
of the universal  mass scale 
$m=2A$ with $M_{1/2}=200$~GeV, $\tan\beta=15$
and order one $\tilde A_{ij}^{u,d}$ generated randomly.
Each panel shows both the LR and RL mass insertions.
The experimental limit is represented  by the curve.
}
\label{fig:lrydg2}
\end{figure}

\begin{figure}
\epsfig{figure=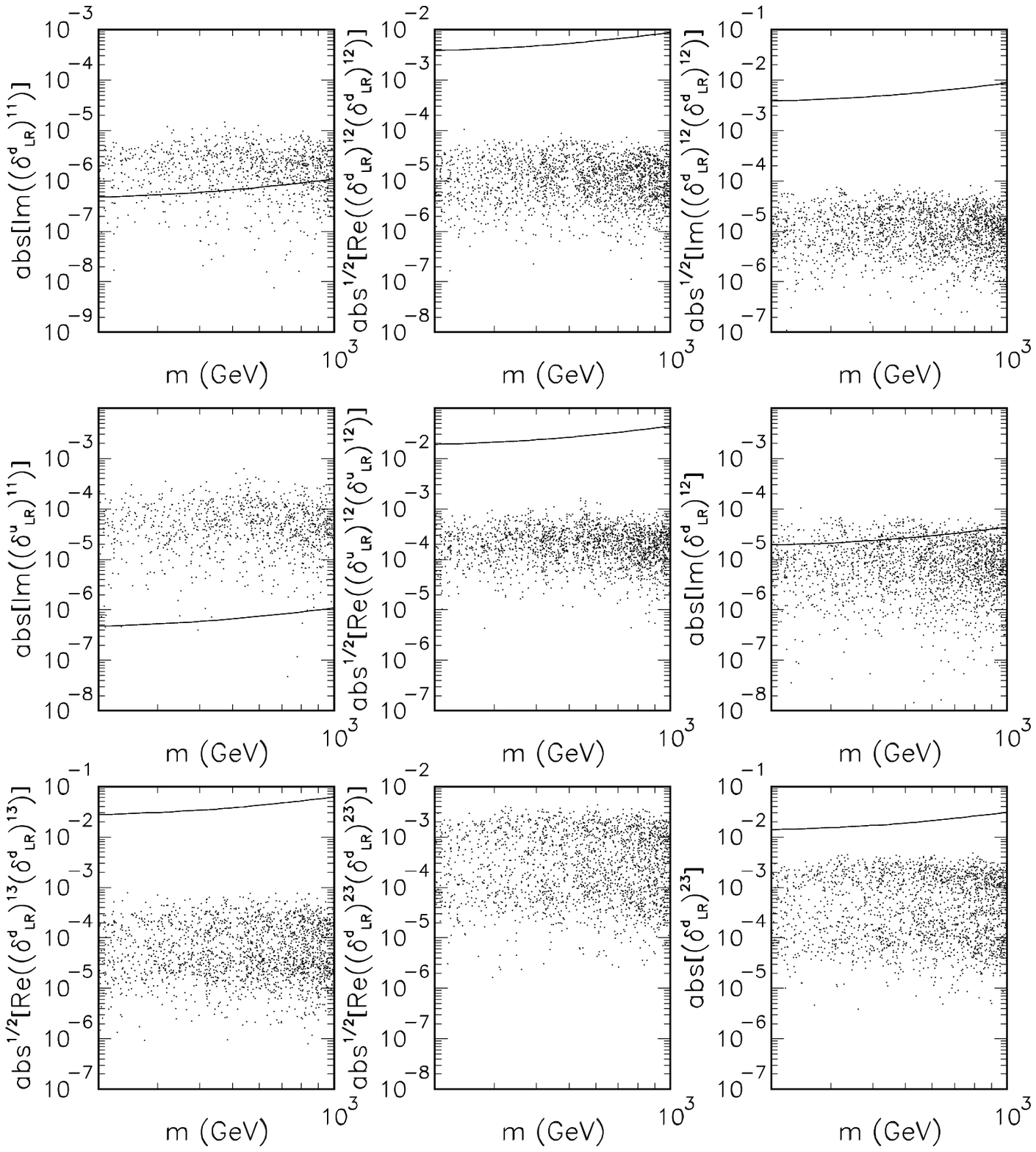,width=\linewidth} 
\vspace{1.0truecm}
\caption{As in Figure \ref{fig:lrydg2}  but
for the hierarchical  Yukawa texture (ii).
}
\label{fig:lr320422g2}
\end{figure}

\begin{figure}
\epsfig{figure=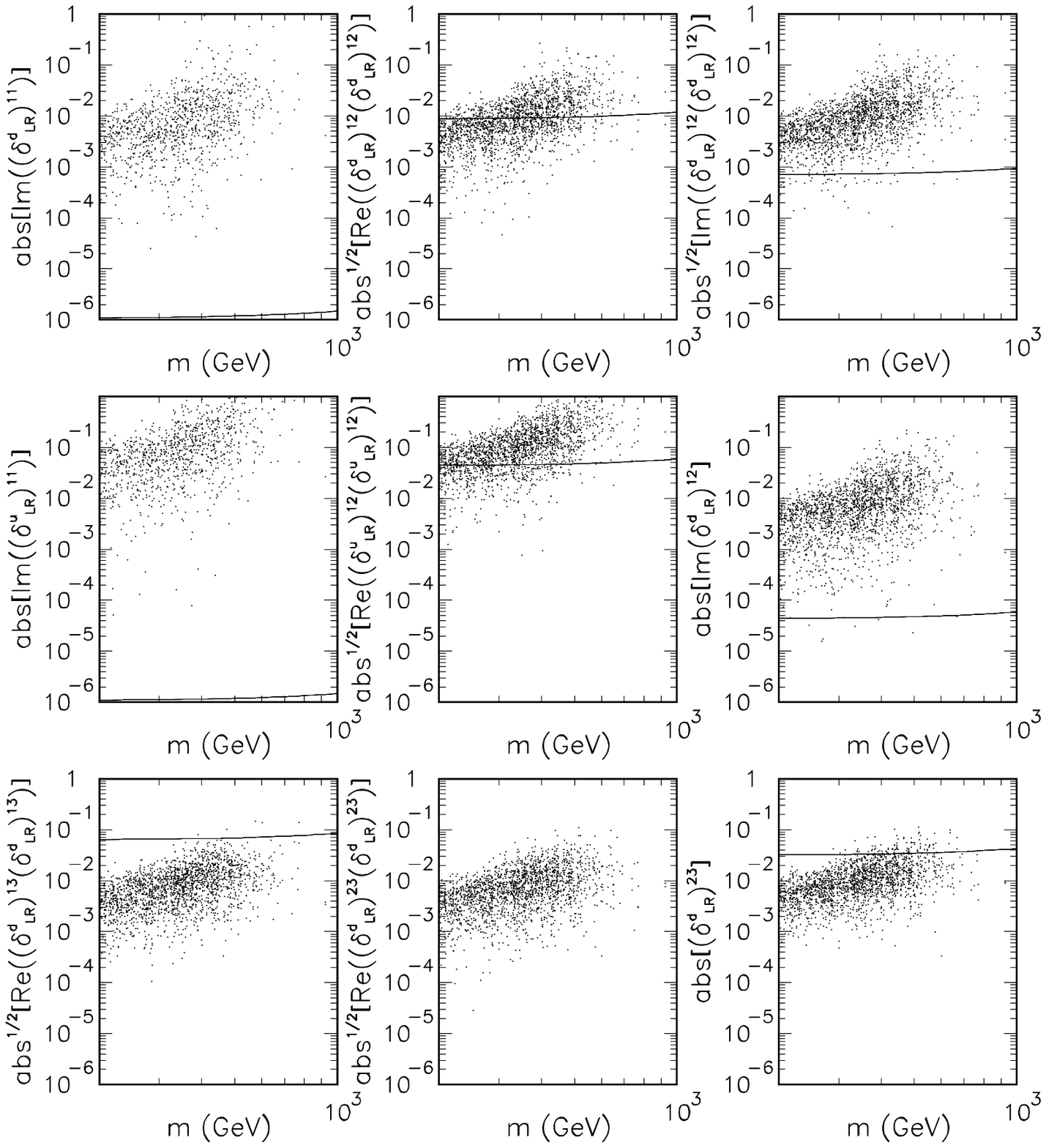,width=\linewidth} 
\vspace{1.0truecm}
\caption{As in Figure \ref{fig:lrydg2} but
for the democratic Yukawa texture (iii) and 
$M_{1/2}=500$~GeV.}
\label{fig:lrdemg5}
\end{figure}

The most important effects are in the LR/RL sector.
Figures~(\ref{fig:lrydg2})-(\ref{fig:lrdemg5})
display the relevant mass insertions and the experimental bounds for  Yukawa 
textures (i)-(iii), respectively. At the GUT scale the gaugino masses are 
fixed to 200 GeV for  textures (i), (ii), and to 500 GeV for  texture 
(iii). The horizontal axis corresponds to the sfermion masses at the GUT scale:
$m_Q^2=m_U^2=m_D^2\equiv m^2$. The mass scale of the $A-$terms is taken
as $\tilde A=m/2$ and  order one non--universal entries of 
$\tilde A_{ij}$ are generated randomly. 
Each panel shows combinations of both the LR and the RL mass insertions, e.g.
$|{\rm Re}((\delta^u_{LR})_{12}(\delta^u_{LR})_{12})|^{1/2}$
and $|{\rm Re}((\delta^u_{RL})_{12}(\delta^u_{RL})_{12})|^{1/2}$. This explains
the presence of two distinct  bands in some of the panels.

We see that for the hierarchical Yukawa textures (i) and (ii), the only 
problematic observables are the EDMs, representing the SUSY CP problem. 
The constraint stemming from $\epsilon^\prime$ is rather mild. The situation 
is much worse for the democratic texture (iii), for which all the limits 
imposed by the Kaon observables are exceeded.
\vskip0.3cm

\noindent{\it Texture (C)}

\begin{figure}
\epsfig{figure=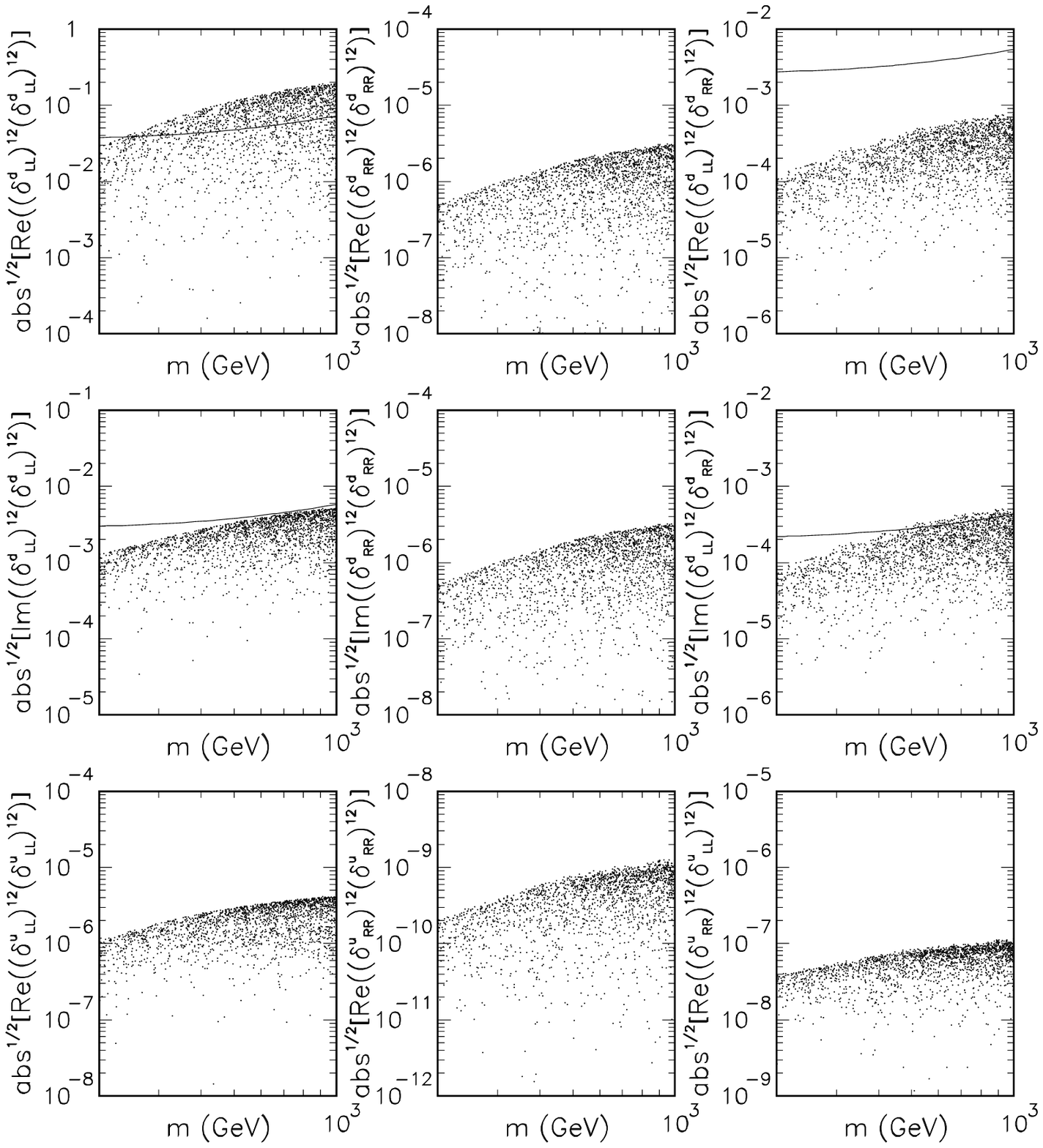,width=\linewidth} 
\vspace{1.0truecm}
\caption{Phenomenologically relevant combinations of the insertions 
$(\delta^{d,u}_{LL})^{12}$ and  $(\delta^{d,u}_{RR})^{12}$
for the hierarchical Yukawa texture (i)
as a function of the universal  mass scale 
$m\equiv m_2=m_3$ with $M_{1/2}=200$~GeV, $A=0$, $\tan\beta=15$
and $m_1$ varied randomly in the range $m/2 \div m$.
The experimental limit is represented  by the curve.
}
\label{fig:yd1kkg2}
\end{figure}

\begin{figure}
\epsfig{figure=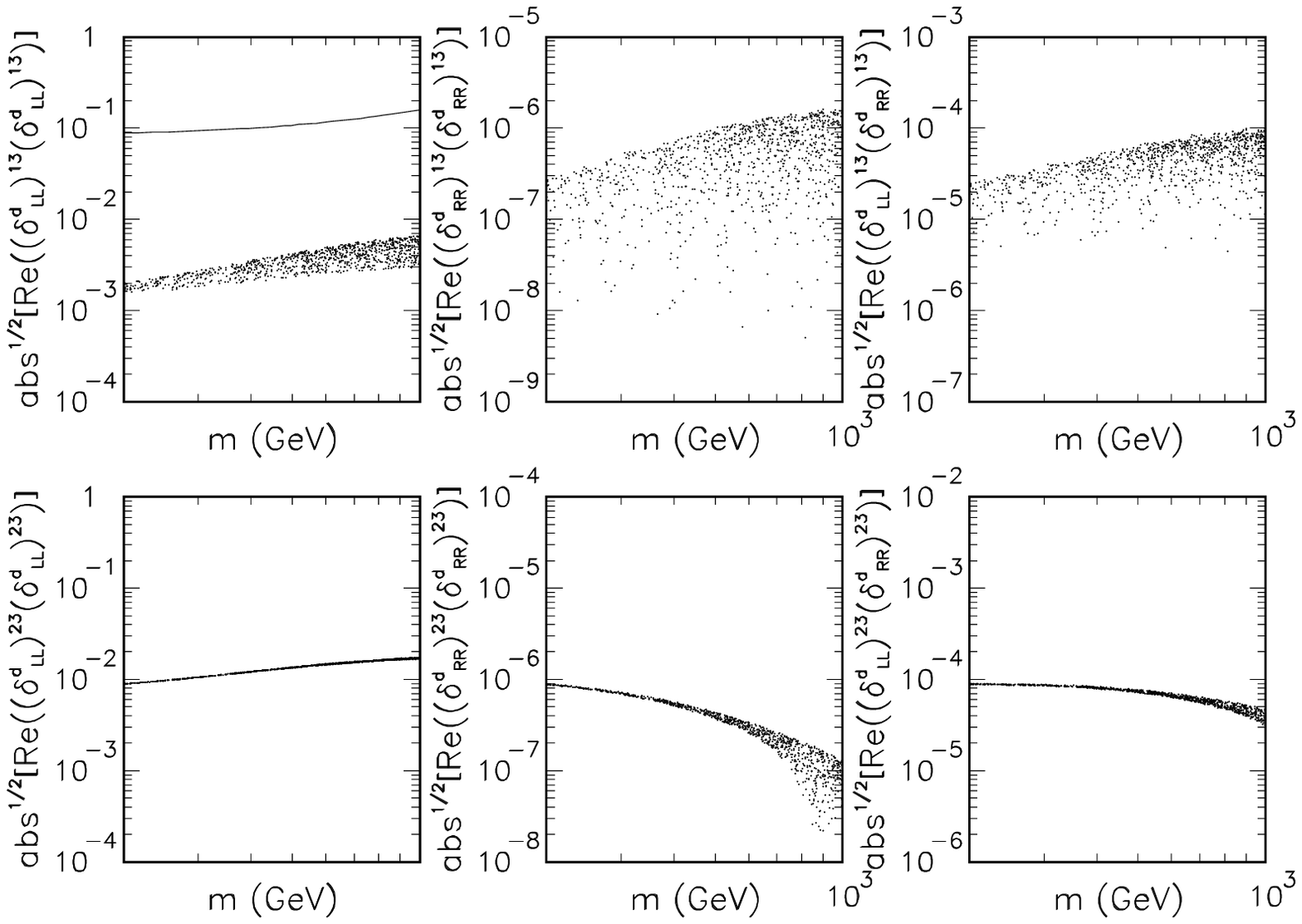,width=\linewidth} 
\vspace{1.0truecm}
\caption{Phenomenologically relevant combinations of the insertions 
$(\delta^d_{LL})^{13}$, $(\delta^d_{RR})^{13}$, $(\delta^d_{LL})^{23}$, 
$(\delta^d_{RR})^{23}$
for the hierarchical Yukawa texture (i)
as a function of the universal  mass scale 
$m\equiv m_2=m_3$ with $M_{1/2}=200$~GeV, $A=0$, $\tan\beta=15$
and $m_1$ varied randomly in the range $m/2 \div m$.
The experimental limit is represented  by the curve.
}
\label{fig:yd1bbg2}
\end{figure}

\begin{figure}
\epsfig{figure=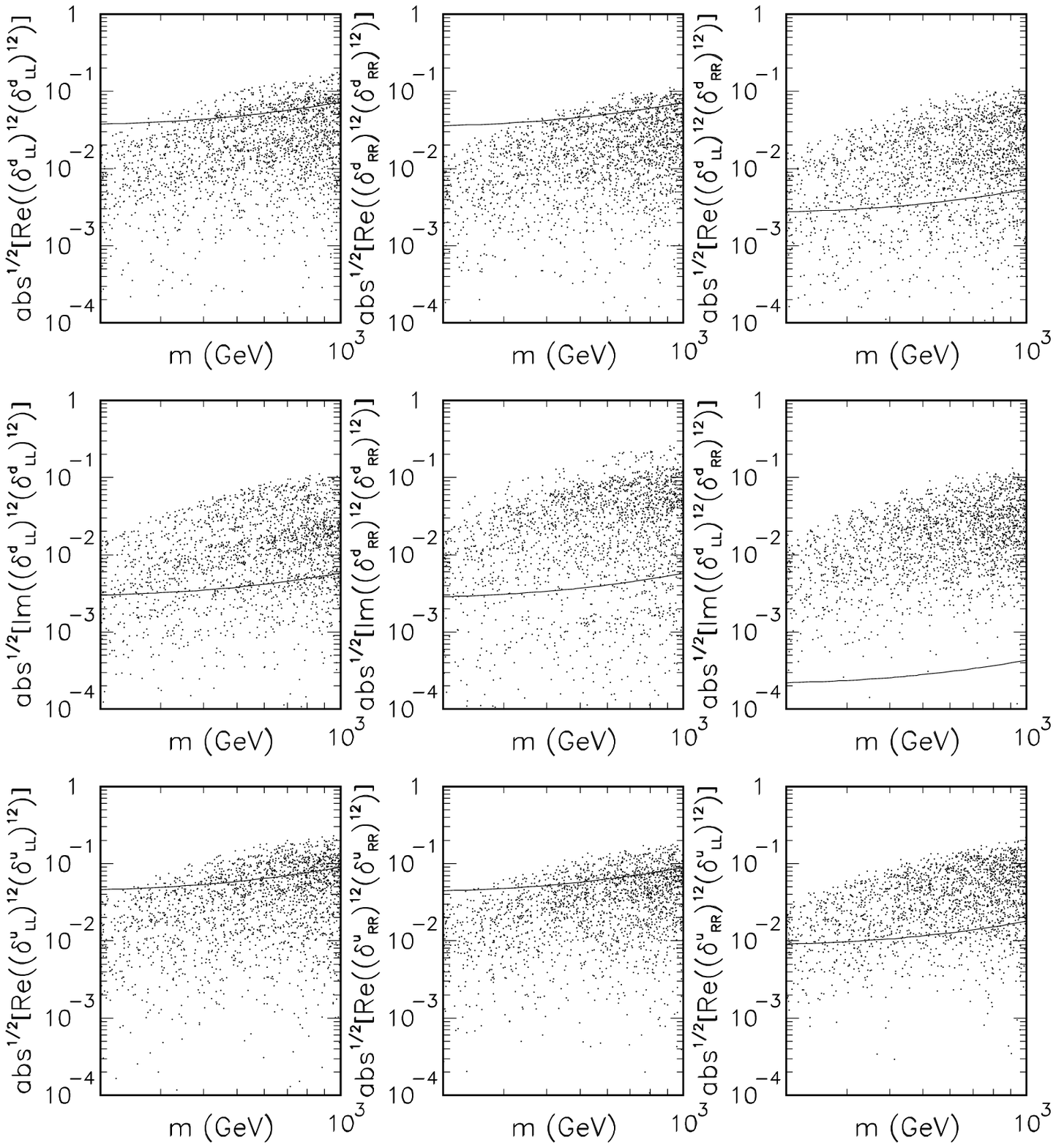,width=\linewidth} 
\vspace{1.0truecm}
\caption{As in Figure \ref{fig:yd1kkg2} but 
for the hierarchical  Yukawa texture (ii).
}
\label{fig:lav1kkg2}
\end{figure}

\begin{figure}
\epsfig{figure=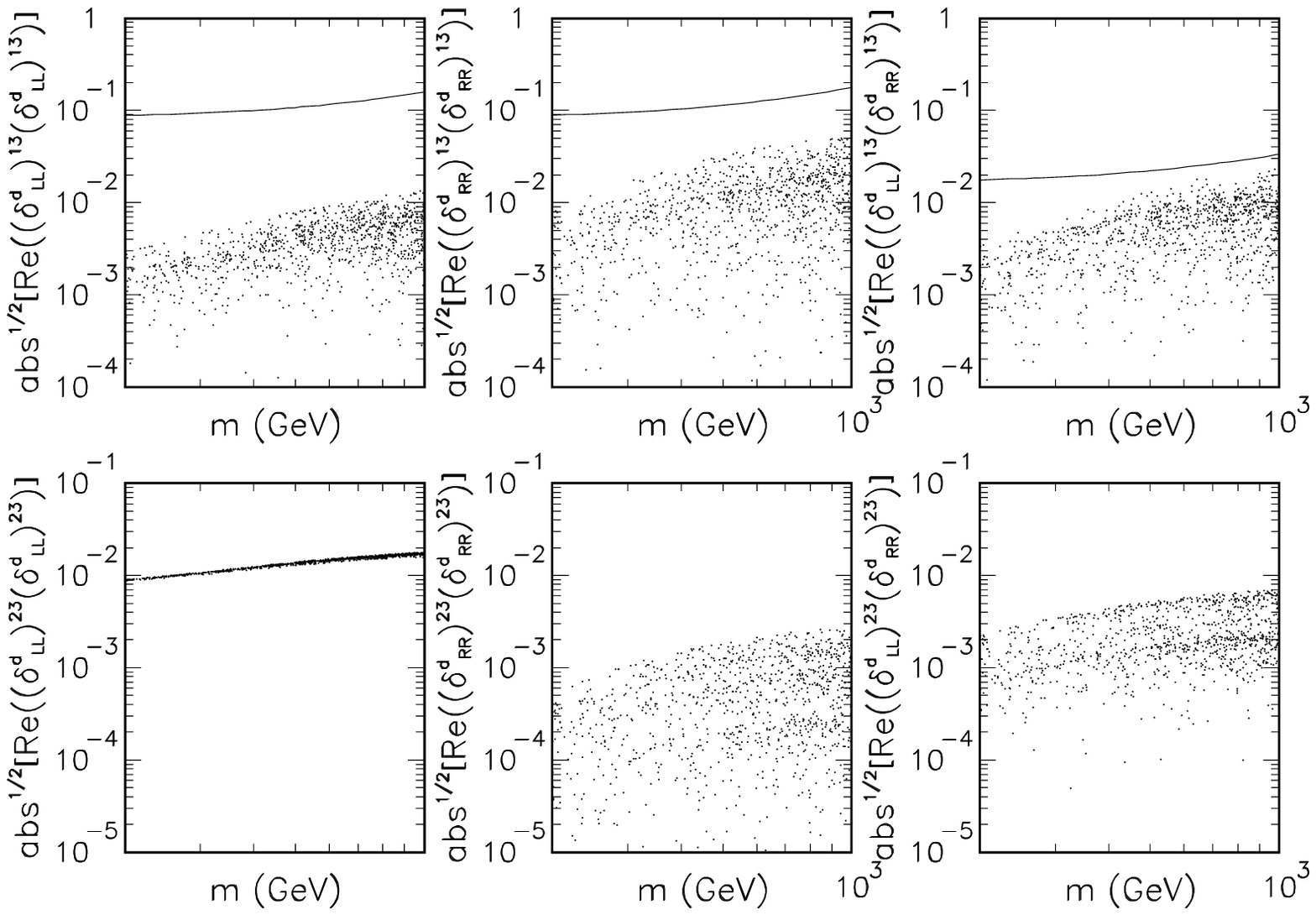,width=\linewidth} 
\vspace{1.0truecm}
\caption{As in Figure \ref{fig:yd1bbg2} but 
for the hierarchical  Yukawa texture (ii).
}
\label{fig:lav1bbg2}
\end{figure}

\begin{figure}
\epsfig{figure=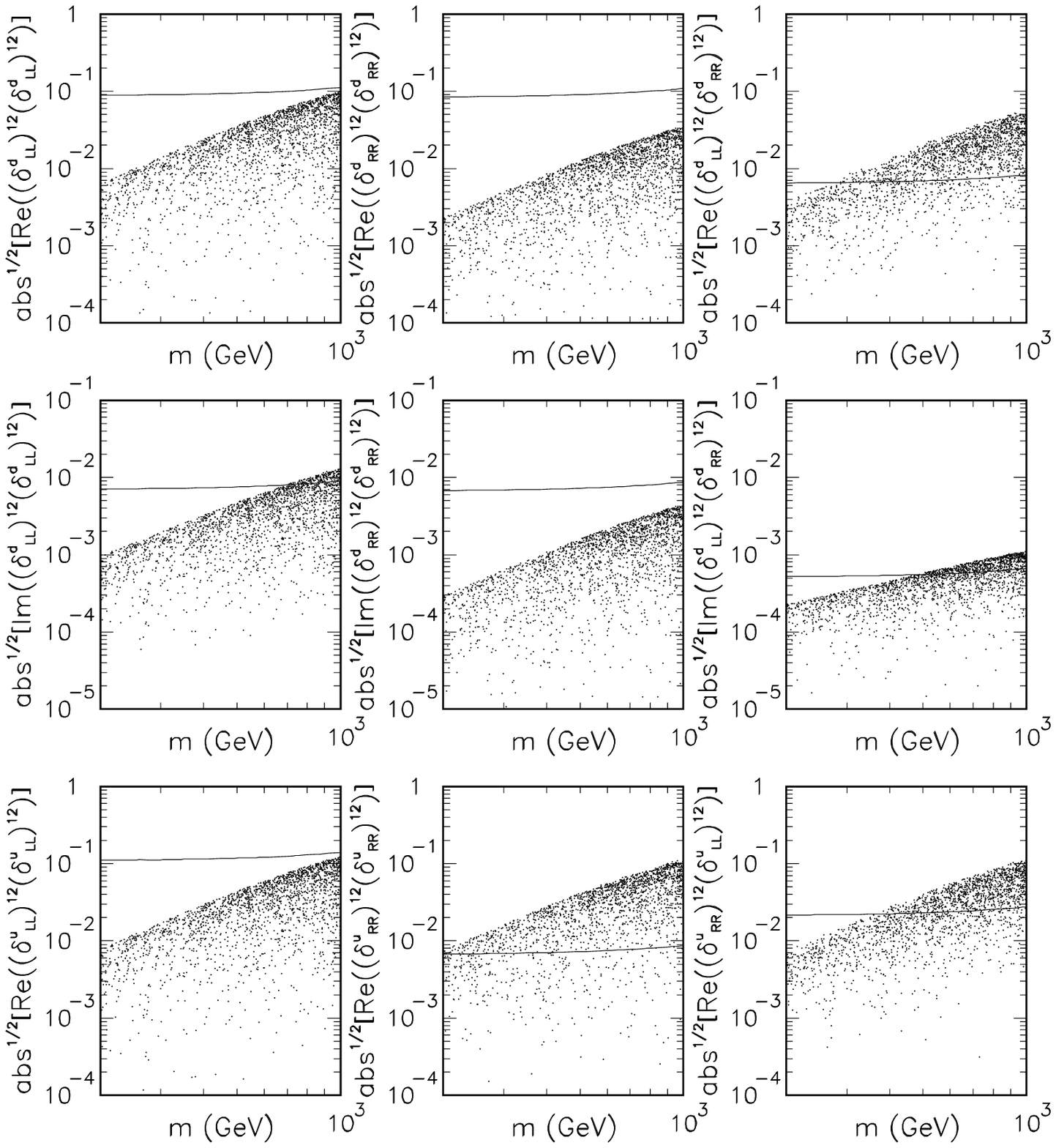,width=\linewidth} 
\vspace{1.0truecm}
\caption{As in Figure \ref{fig:yd1kkg2} but 
for the democratic Yukawa texture (iii) and $M_{1/2}=500$~GeV.}
\label{fig:dem1kkg5}
\end{figure}

\begin{figure}
\epsfig{figure=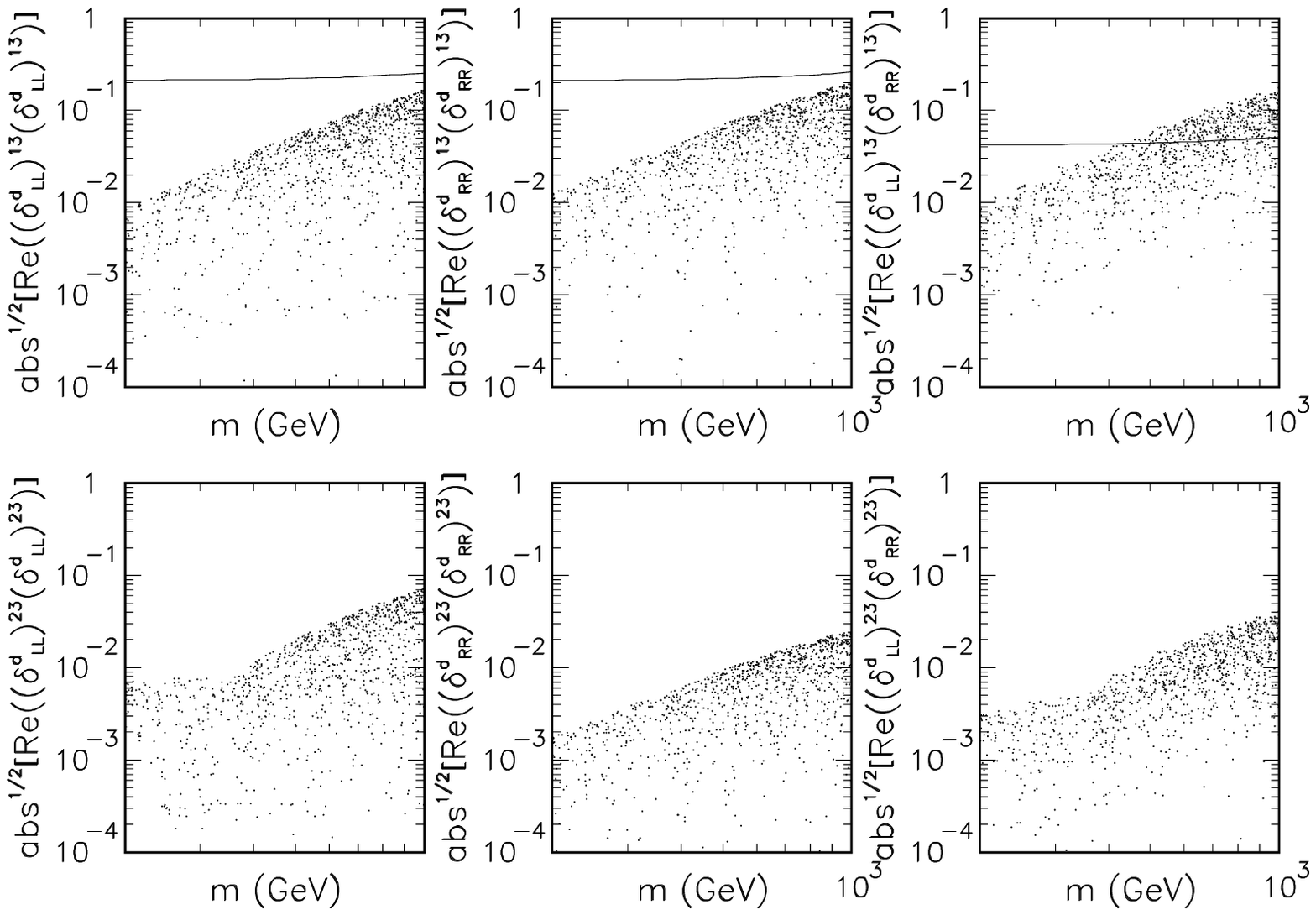,width=\linewidth} 
\vspace{1.0truecm}
\caption{As in Figure  \ref{fig:yd1bbg2}  but 
for the democratic Yukawa texture (iii) and $M_{1/2}=500$~GeV.}
\label{fig:dem1bbg5}
\end{figure}

Significant LL and RR mass insertions are induced. In 
Figures~(\ref{fig:yd1kkg2})-(\ref{fig:dem1bbg5}),
relevant mass insertions are shown for the GUT scale boundary condition
\begin{equation}
m^2_Q=m^2_U=m^2_D= \left(\matrix{m^2_1 &   0    & 0    \cr
                                 0     &  m^2 & 0    \cr
                                 0     &   0    & m^2}\right) 
\end{equation}
and Yukawa textures (i)-(iii). As before, the gaugino masses are fixed to 
200 GeV for  textures (i), (ii), and to 500 GeV for  texture (iii). 
$m_1$ is generated randomly in the range 
${1\over2}m\div m$. 
(In fact, we allow for larger departures from universality 
than usually exists in typical semirealistic models, see Section 2).

Figures~(\ref{fig:yd1kkg2})-(\ref{fig:dem1bbg5}) show 
 that the mass insertions grow 
with $m$, which is perhaps counterintuitive.
The reason for that  is the gluino loop renormalization effect which
is more important for small values of $m$ \cite{Choudhury:1994pn}. The 
low--energy degeneracy parameter $(m^2-m^2_1)/(m^2 + \Delta_{\tilde g} m^2)$, 
where $\Delta_{\tilde g} m^2$ is induced by the RG running, grows with $m^2$ 
leading to larger mass insertions. This ``aligning'' gluino  effect is very  
important and can reduce a  mass insertion by up to an order of
magnitude for similar squark and gluino masses. For larger gluino masses it 
is even more important.

For  $m \sim M_{1/2}$ and small rotation angles (texture (i)), we see that $no$ 
significant FCNC problem exists. Even if the mass splitting at the GUT scale 
is of order one, $\delta_{LL,RR}^{12}$ are suppressed by both the gluino loop
RG effect and  the small rotation angle (see Eq.~(\ref{deg})).

For large rotation angles in the right-handed sector (texture (ii)), the 
problem of FCNC becomes more acute, mainly due to the simultaneous presence of 
large LL and RR mass insertions. The most constraining observable here
is $\epsilon$. The problem  disappears eventually as $M_{1/2}$ increases,
yet it still persists for $M_{1/2}=500$ GeV. This is also true for 
texture (iii) as seen in Figs.~(\ref{fig:dem1kkg5},\ref{fig:dem1bbg5}).

An interesting ``focusing'' effect is seen in Figures (\ref{fig:yd1bbg2}) and
(\ref{fig:lav1bbg2}). The values of $(\delta^d_{23})_{LL}$ are independent of
$m_1$ and of the specific Yukawa texture as long as it is diagonalized by a small
angle rotation in the left--handed sector. This is because the  dominant
contribution to $(\delta^d_{23})_{LL}$ is due to the top Yukawa RG effect
which depends on the CKM matrix only.

\begin{figure}
\epsfig{figure=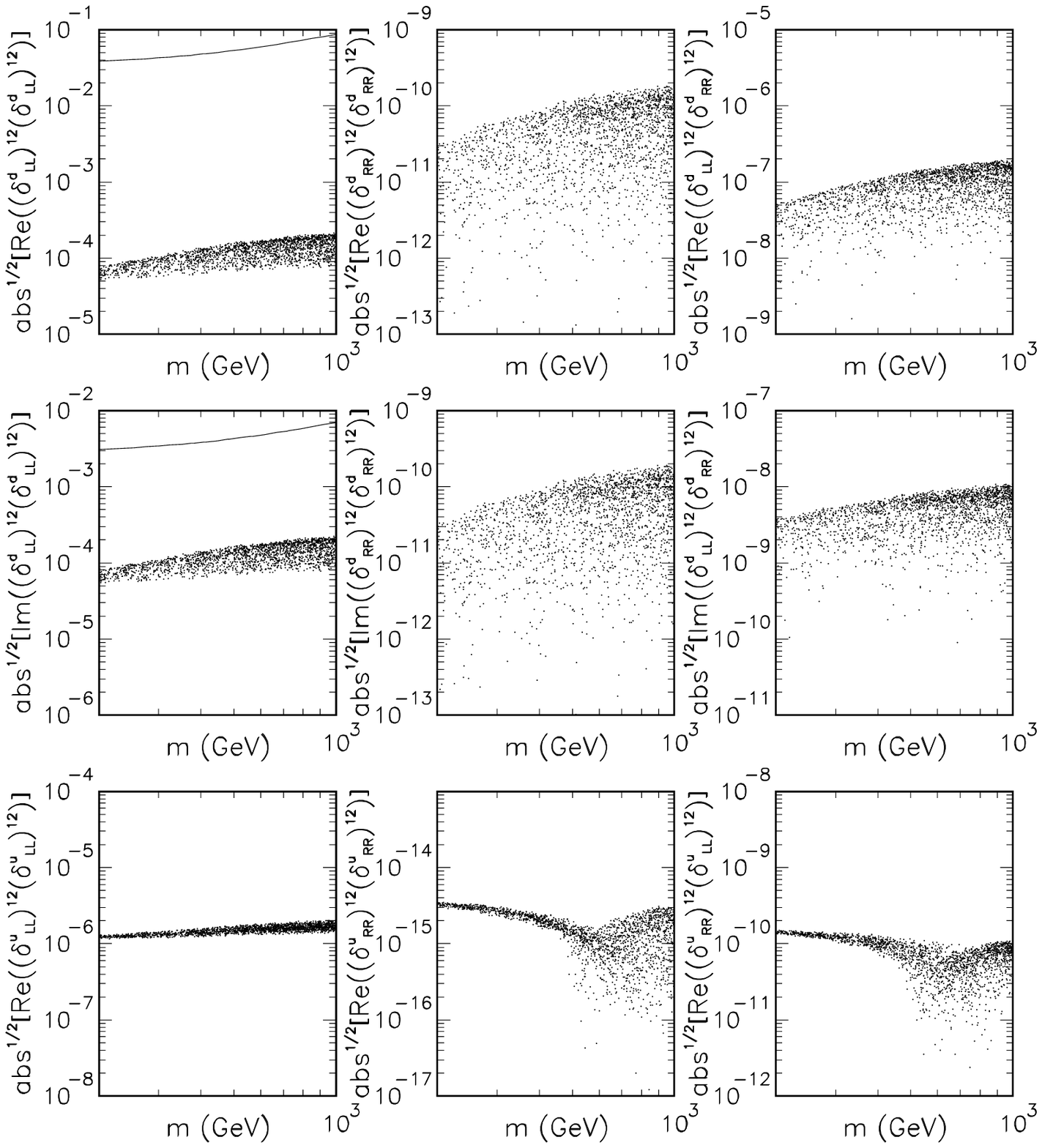,width=\linewidth} 
\vspace{1.0truecm}
\caption{Phenomenologically relevant combinations of the insertions 
$(\delta^{d,u}_{LL})^{12}$ and  $(\delta^{d,u}_{RR})^{12}$
for the hierarchical Yukawa texture (i)
as a function of the universal  mass scale 
$m\equiv m_1=m_2$ with $M_{1/2}=200$~GeV, $A=0$, $\tan\beta=15$
and $m_3$ varied randomly in the range $m/2 \div m$.
The experimental limit is represented  by the curve.
}
\label{fig:yd3kkg2}
\end{figure}

\begin{figure}
\epsfig{figure=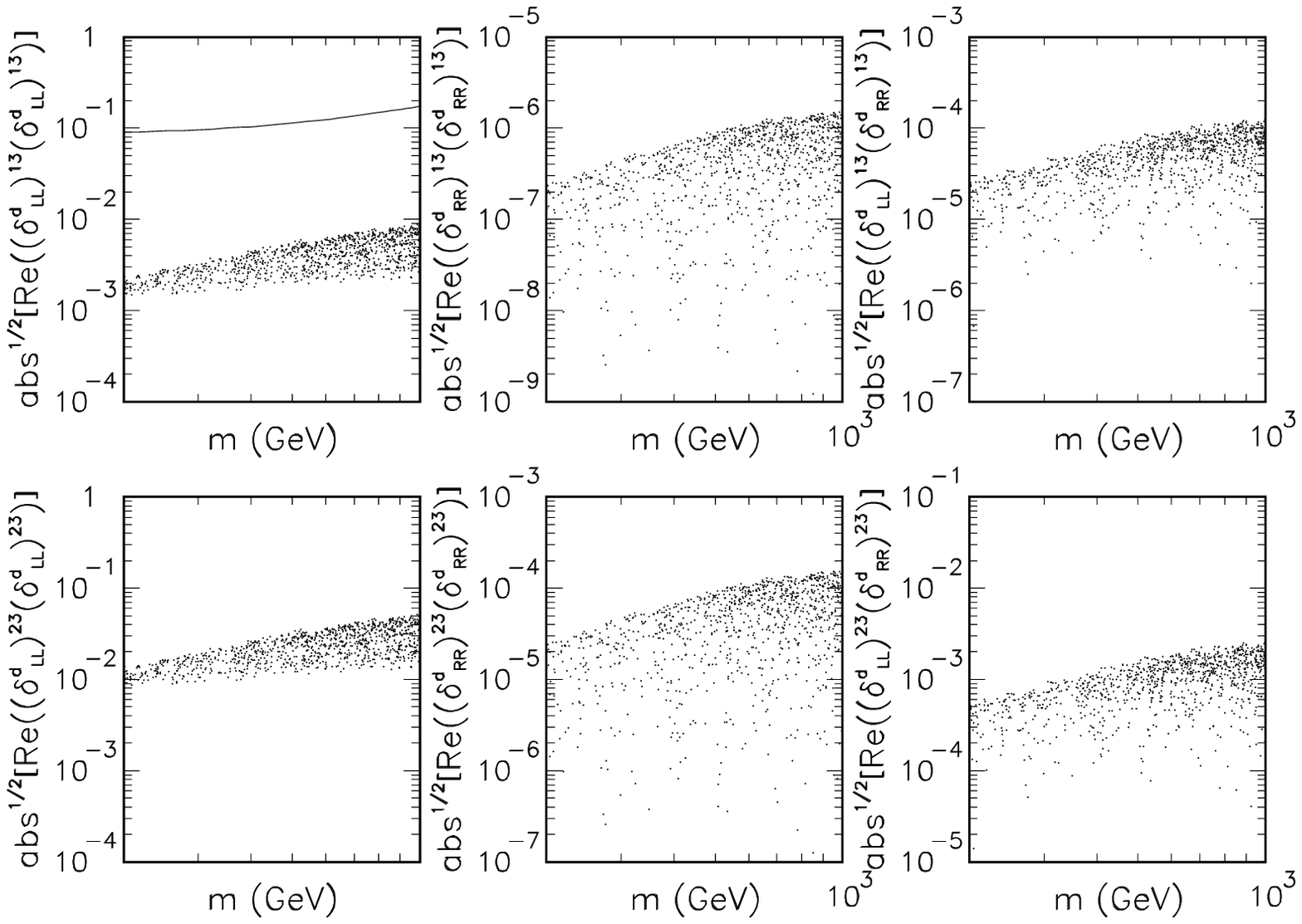,width=\linewidth} 
\vspace{1.0truecm}
\caption{Phenomenologically relevant combinations of the insertions 
$(\delta^d_{LL})^{13}$, $(\delta^d_{RR})^{13}$, $(\delta^d_{LL})^{23}$, 
$(\delta^d_{RR})^{23}$
for the hierarchical Yukawa texture (i)
as a function of the universal  mass scale 
$m\equiv m_1=m_2$ with $M_{1/2}=200$~GeV, $A=0$, $\tan\beta=15$
and $m_3$ varied randomly in the range $m/2 \div m$.
The experimental limit is represented  by the curve.
}
\label{fig:yd3bbg2}
\end{figure}

\begin{figure}
\epsfig{figure=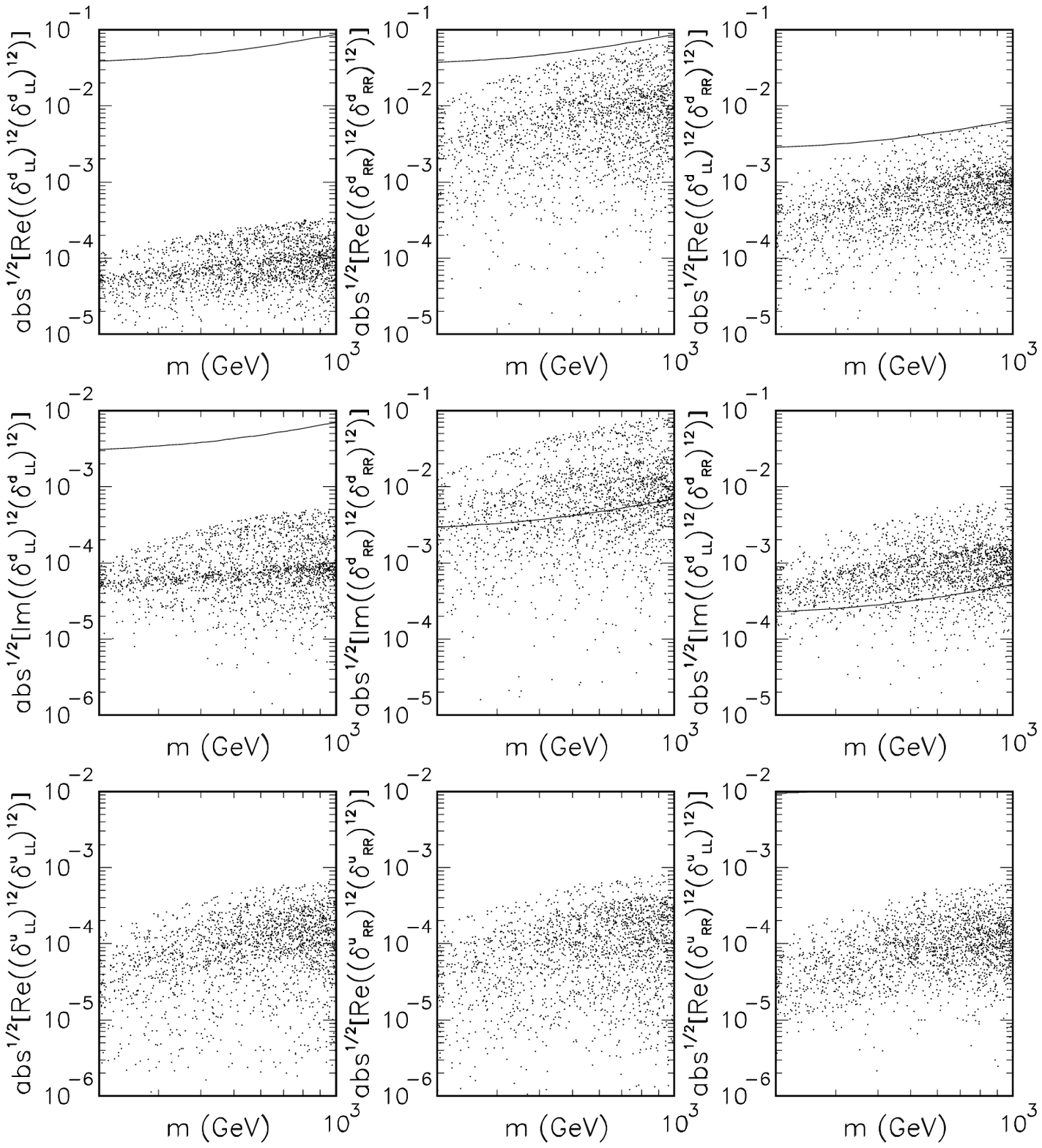,width=\linewidth} 
\vspace{1.0truecm}
\caption{
As in Figure \ref{fig:yd3kkg2} but 
for the hierarchical  Yukawa texture (ii).
}
\label{fig:lav3kkg2}
\end{figure}

\begin{figure}
\epsfig{figure=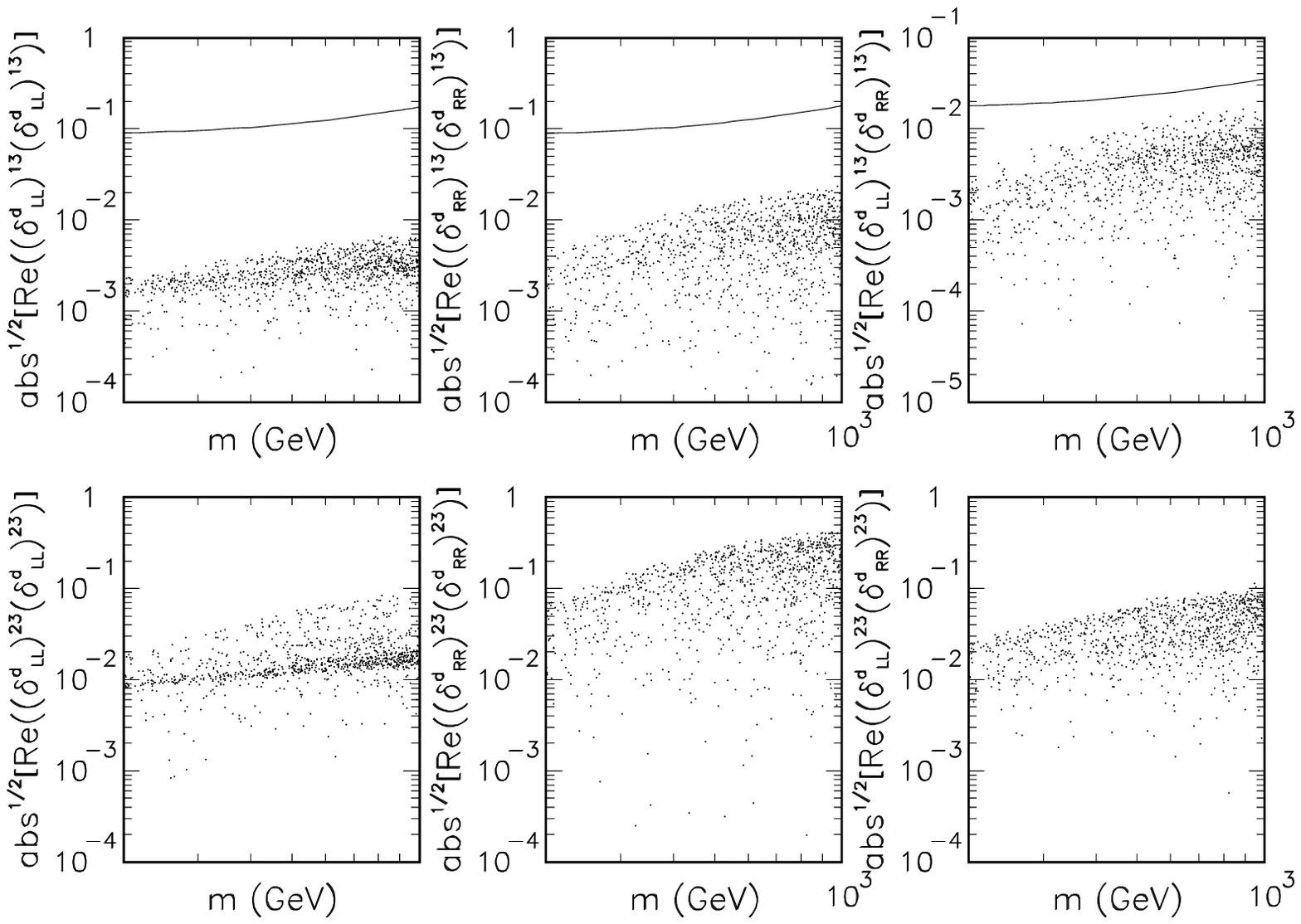,width=\linewidth} 
\vspace{1.0truecm}
\caption{
As in Figure \ref{fig:yd3bbg2} but 
for the hierarchical  Yukawa texture (ii).
}
\label{fig:lav3bbg2}
\end{figure}

\begin{figure}
\epsfig{figure=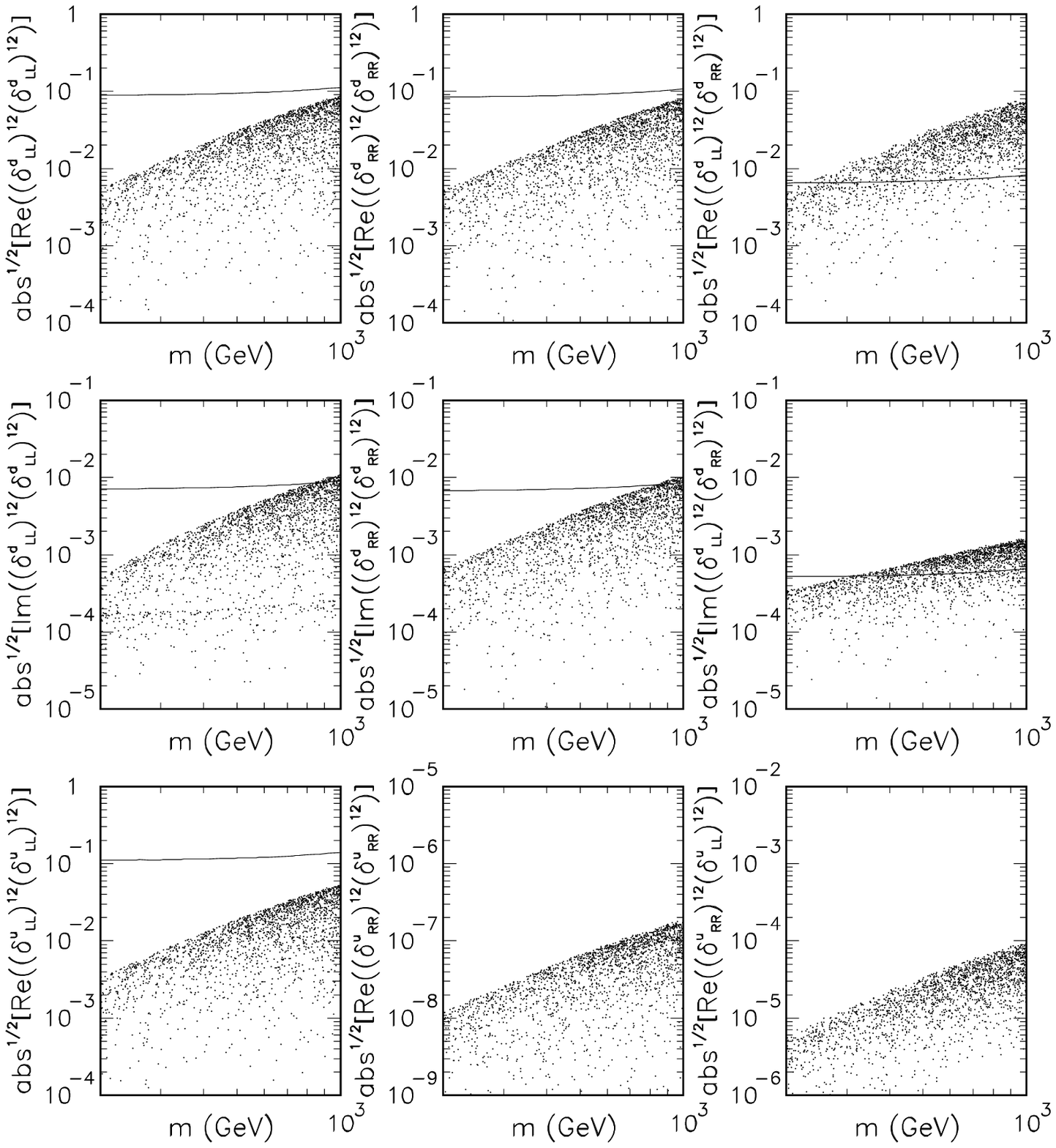,width=\linewidth} 
\vspace{1.0truecm}
\caption{As in Figure  \ref{fig:yd3kkg2} but 
for the democratic Yukawa texture (iii) and $M_{1/2}=500$~GeV.}
\label{fig:dem3kkg5}
\end{figure}

\begin{figure}
\epsfig{figure=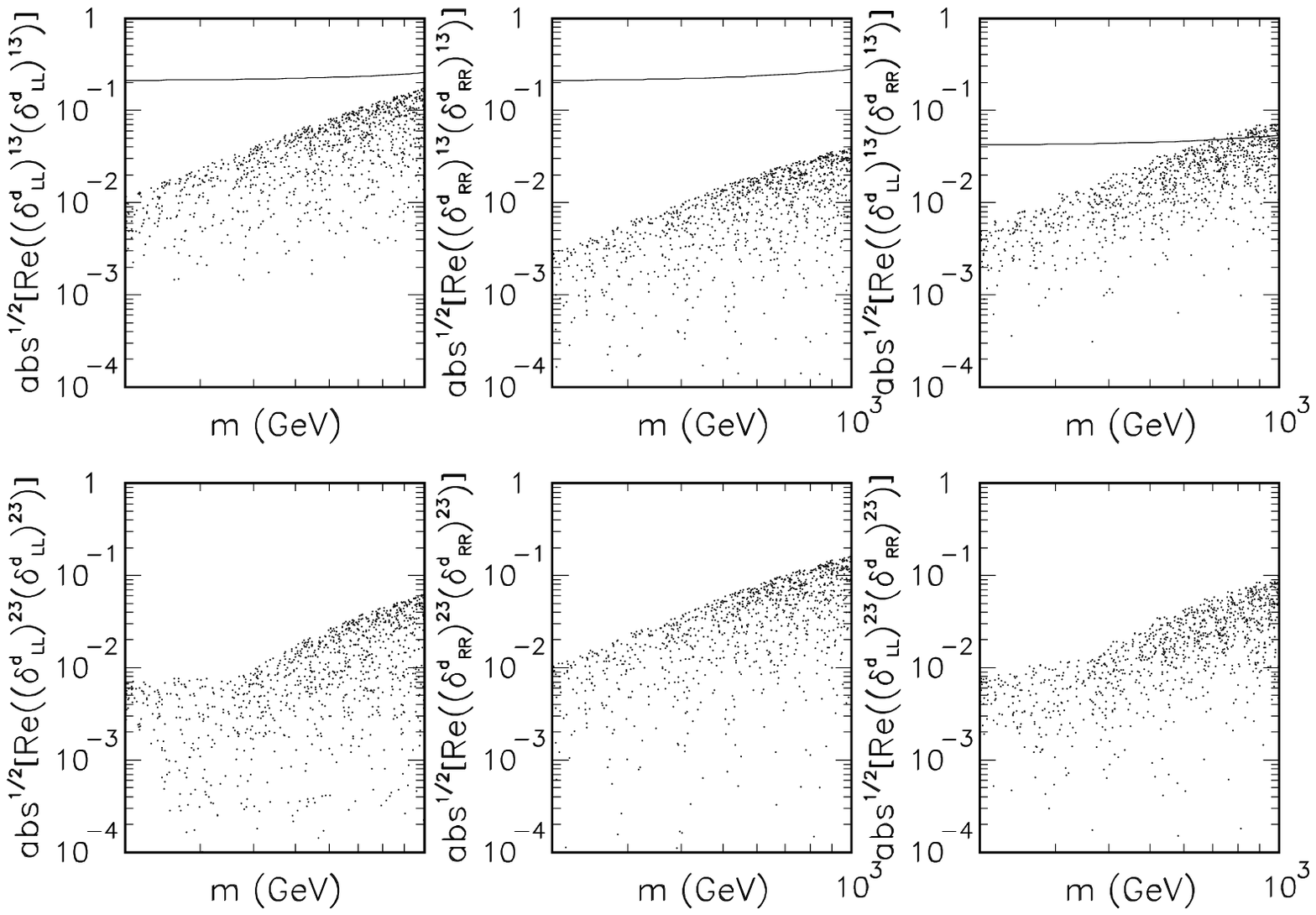,width=\linewidth} 
\vspace{1.0truecm}
\caption{
As in Figure \ref{fig:yd3bbg2}  but 
for the democratic Yukawa texture (iii) and $M_{1/2}=500$~GeV.
}
\label{fig:dem3bbg5}
\end{figure}

Figures~(\ref{fig:yd3kkg2})-(\ref{fig:dem3bbg5}) show our 
results for the case of degenerate first two generations,
\begin{equation}
m^2_Q=m^2_U=m^2_D= \left(\matrix{m^2   &   0  & 0    \cr
                                 0     &  m^2 & 0    \cr
                                 0     &   0  & m^2_3}\right) \;,
\end{equation}
with $m_3$ generated randomly in the range ${1\over2}m\div m$.
Again, no problem with FCNC arises for texture (i). For texture (ii), there 
is some tension with $\epsilon$ due to large rotations in the right--handed 
sector. The problem becomes milder with increasing gluino mass and 
dissappears for $M_{1/2}=500$ GeV. 
The FCNC problem is serious for the 
Yukawa texture (iii), in which case the limits stemming from the Kaon and 
$D-$meson observables are exceeded for $M_{1/2}=200$ GeV
(the $B^0_d$-$\bar B^0_d$ mixing imposes
only a mild constraint).
However, no significant FCNC problem exists
for heavier gaugino masses, $M_{1/2}=500$ GeV,
as shown in Figures
 (\ref{fig:dem3kkg5}) and (\ref{fig:dem3bbg5}).

For texture (D), similar conclusions can be drawn by combining the results for 
textures (B) and (C). The main point is that the $A-$term nonuniversality is 
essentially unconstrained (ignoring the CP problem), while some degeneracy of 
the diagonal soft masses may be required.

\begin{table}[h]
\begin{center}
\begin{tabular}{|c||c|c|}
\hline
Observable & K\"ahler potential flavour violation & $A-$term flavour
violation  \\
\hline
$\Delta M_K$ &  problem &  no problem  \\
$\epsilon$ &  problem &  no problem   \\
$\epsilon'$ &  no problem &  no problem \\
BR($K \rightarrow \pi \nu \bar \nu$) &  no problem &  no problem \\
$\Delta M_D$ &  problem &  no problem \\
$\Delta M_{B_d}$  &  problem &  no problem\\ 
BR($ b \rightarrow s \gamma$)  &  no problem &  no problem \\
$A_{\rm CP}(B\rightarrow \psi K_s)$ & problem & no problem \\
\hline
\end{tabular}
\end{center}
\caption{
Observables and their sensitivity to the source of flavour violation.
Here ``K\"ahler potential flavour violation'' refers to misalignment
between the K\"ahler potential and the soft scalar masses of 
Eq.~(\ref{eqn:mA}) (texture (E)),
while ``$A-$term flavour violation'' refers to misalignment between the
$A-$terms and the Yukawa matrices. The entries indicate whether order one 
non-universality at the high energy scale is in conflict with the 
particular observable when hierarchical Yukawa matrices are assumed.
\label{table1}}
\end{table}

Finally, we compare in Table 2 SUSY flavour violation 
resulting from misalignment between the K\"ahler potential and the soft scalar 
masses (texture (E)) with flavour violation resulting from misalignment  
between the $A-$terms and the Yukawa matrices. Clearly, the former scenario
is strongly constrained. This implies that the K\"ahler potential and the 
soft scalar mass terms are diagonal (to a good approximation) in the same 
basis. Generally, there are further constraints on the diagonal entries of the 
soft mass squared matrices. These constraints strongly depend on the Yukawa 
textures. If the  Yukawa matrices are diagonalized by large angle rotations, the 
diagonal entries must be degenerate to a large extent. On the other hand, if 
the Yukawa matrices are diagonalized by rotation matrices similar to the CKM one,
order one splittings among the diagonal entries are allowed.

\section{Comments on the lepton sector}

So far we have been focusing on the squark sector. In the lepton sector, the 
analysis becomes more involved due to unknown origin and nature of neutrino 
masses. For example, if they originate from the seesaw mechanism operating
at some high scale $M_R<M_{\rm string}$, one should also take into account 
effects of the additional Yukawa couplings which generate off--diagonal
entries in the slepton mass matrices during the RG evolution. We do not
undertake such an analysis here. Instead, 
we only make  some qualitative remarks (neglecting the RG 
effects).

The most restrictive observables are the 
$l_i\rightarrow l_j\gamma$ branching ratios \cite{Lavignac:2003tk}:
\begin{eqnarray}
&& {\rm BR}(\mu \rightarrow e \gamma) <1.2 \times 10^{-11} \;, \nonumber\\
&& {\rm BR}(\tau \rightarrow e \gamma) <2.7 \times 10^{-6} \;, \nonumber\\
&& {\rm BR}(\tau \rightarrow \mu \gamma) <5.0 \times 10^{-7} \;.
\end{eqnarray} 
The corresponding  constraints on the mass insertions can be obtained by 
rescaling the original results of Ref.\cite{Gabbiani:1996hi},
\begin{eqnarray}
&&\Bigl\vert\left(\delta^l_{12}\right)_{LL}\Bigr\vert
<4\times10^{-3}~,\phantom{aaa}
\Bigl\vert\left(\delta^l_{12}\right)_{LR}\Bigr\vert
<1\times10^{-6}~,\nonumber\\
&&\Bigl\vert\left(\delta^l_{13}\right)_{LR}\Bigr\vert
<2\times10^{-2}~,\phantom{aaa}
\Bigl\vert\left(\delta^l_{23}\right)_{LR}\Bigr\vert
<1\times10^{-2} \label{eqn:lepton_bounds}
\end{eqnarray}
for the photino and slepton masses of 100 GeV. The constraint on   
$\left(\delta^l_{12}\right)_{RR}$ is the same as that on 
$\left(\delta^l_{12}\right)_{LL}$, while the other insertions are 
essentially unconstrained.

The bounds (\ref{eqn:lepton_bounds}) immediately tell us  that the LL and 
RR blocks of the slepton 
mass matrix in the (12) sector are  proportional to the unit matrix with 
good accuracy.
As to the LR sector, a simple estimate of $\left(\delta^l_{12}\right)_{LR}$ gives
\begin{equation}
(\delta_{12}^l)_{LR} \sim \alpha_{12}{m_e\over\tilde M} +
\beta_{12}{m_\mu\over\tilde M} + \gamma_{12}{m_\tau\over\tilde M} \;
\end{equation}
with $\alpha_{12}$, $\beta_{12}$, $\gamma_{12}\simlt{\cal O}(1)$. 
The bounds then imply 
\begin{equation}
\alpha_{12}\simlt 10^{-1}~,\phantom{aa}\beta_{12}\simlt10^{-3}~,
\phantom{aa}\gamma_{12}\simlt 10^{-4}\;.
\end{equation}
Thus, the alignment between the $A-$terms and the lepton Yukawa matrices has
to be very precise,\footnote{We note that the constraints on
$\left(\delta^l_{13} \right)_{LR}$ and $\left(\delta^l_{23} \right)_{LR}$ are 
trivially satisfied since their magnitude is bounded by $m_\tau/\tilde m$.}
in sharp contrast with the squark sector. 
This suggests that the 
charged lepton Yukawa matrix is  diagonal at the high energy scale, 
presumably due to some symmetry such as the lepton number, or that the 
$A-$terms are universal.

The bound on ${\rm BR}(\mu \rightarrow e \gamma) $ is expected to be improved
by 3 orders of magnitude at PSI. The above estimates suggest that a non--zero 
signal is expected. A negative result would mean that the Yukawa matrices and 
the $A-$terms are extremely well aligned indicating a special symmetry or a 
special nature of the SUSY breaking.

\section{Prospects}

Different processes are sensitive to different chirality types of mass 
insertions, making it possible (at least in principle) to distinguish 
various supergravity textures. Below we give a few examples supporting 
this point.

The estimates (\ref{deltai3}) mean that SUSY flavour effects grow with quark 
masses. In particular,  the magnitude of the LR mass insertions involving the 
third generation is enhanced in SUGRA models, which renders their effect 
potentially observable. They can significantly affect processes sensitive to 
LR mass insertions. Such processes are loop suppressed in the Standard Model 
and are usually due to penguin--type diagrams. Apart from $\epsilon^\prime$ 
and Br$(\bar B\rightarrow X_s\gamma)$, a good example is the  
$B^0_d\rightarrow\phi K_S$ decay (for a recent discussion, see 
\cite{Kane:2003zi}). In this case, the decay is due to the 
$b\rightarrow s \bar s s$ transition and 
the ratio of the SUSY and SM decay amplitudes  for squark and gluino masses 
of 500 GeV is given by \cite{Khalil:2003bi}  
\begin{equation}
\left({A^{\rm SUSY}\over A^{\rm SM}}\right)_{B\rightarrow \phi K_S} \simeq
100\times[(\delta_{LR}^d)_{23}+(\delta_{RL}^d)_{23}]+
0.2\times[(\delta_{LL}^d)_{23}+(\delta_{RR}^d)_{23}] \;.
\end{equation}
Clearly, for $(\delta_{LR}^d)_{23}\sim 10^{-2}$ consistent with the 
$b\rightarrow s \gamma$ constraint, the SUSY and SM contributions are
of similar magnitude. On the other hand, the chirality conserving insertions 
contribute far less significantly. Since in the SM the CP asymmetry in this 
decay coincides with that in the $B^0_d\rightarrow\psi K_S$ decay, it provides a  
sensitive probe for new physics contributions to the $b\rightarrow s\bar ss$ 
transition and, in particular, for supersymmetric contributions (see e.g. 
\cite{Nir:2002gu}). In the pseudoscalar channel of
the $b\rightarrow s \bar s s$ transition, $B^0_d\rightarrow\eta^\prime K_S$,
the SM predicts the same (within  5-10\%) CP asymmetry, whereas
supersymmetry gives \cite{Khalil:2003bi}
\begin{equation}
\left({A^{\rm SUSY}\over A^{\rm SM}}\right)_{B^0_d\rightarrow\eta^\prime K_S} 
\simeq100\times[(\delta_{LR}^d)_{23}-(\delta_{RL}^d)_{23}]+
0.2\times[(\delta_{LL}^d)_{23}-(\delta_{RR}^d)_{23}] \;.
\end{equation}
Thus, in SUGRA models one expects (correlated) deviations in both decays.

Another well known example of a process sensitive to LR mass insertions is 
the $b\rightarrow s \gamma$ transition (see e.g. \cite{Gabbiani:1996hi}). 
As is clear from the bound (\ref{eqn:bsg}), $(\delta^d_{23})_{LR}$ of order 
$10^{-2}$, which is natural in SUGRA models, is sufficient to produce significant 
deviations from the SM prediction. On the other hand, to have a similar 
effect from the LL sector, a large $(\delta^d_{23})_{LL} = {\cal O}(1)$ would 
be required. We note that the observation of a direct CP asymmetry in the 
$b\rightarrow s \gamma$ transition of order few percent would be a clean 
signal of new physics, since the well controlled SM prediction yields the 
asymmetry of less than one percent.

An example of a process particularly sensitive to LL and RR insertions
is the $B^0_s$-$\bar B^0_s$ mixing. 
For moderate $\tan\beta$,
the SUSY to SM ratio of the mixing amplitudes is 
\cite{Ball:2003se}
\begin{eqnarray}
\left({M^{\rm SUSY}_{12}\over M^{\rm SM}_{12}}\right)_{B^0_s-\bar B^0_s} 
&\simeq& 
1\times\left[(\delta_{LL}^d)_{23}^2+(\delta_{RR}^d)_{23}^2\right] +
30\times\left[(\delta_{LR}^d)_{23}^2+(\delta_{RL}^d)_{23}^2\right] 
\nonumber\\
&-& 45\times\left[(\delta_{LR}^d)_{23}(\delta_{RL}^d)_{23}\right]
-175\times\left[(\delta_{LL}^d)_{23}(\delta_{RR}^d)_{23}\right] 
\end{eqnarray}
for gluino and squark masses of 500 GeV. At large $\tan\beta$,
other contributions become important \cite{Buras:2002wq}.
Given the limits on 
$(\delta_{LR}^d)_{23}$ from $b\rightarrow s\gamma$, the contribution of 
chirality changing mass insertions is negligible, while even a small 
chirality conserving insertions $(\delta_{LL}^d)_{23}$ and 
$(\delta_{RR}^d)_{23}$ of order ${\cal O}(10^{-1})$ can induce a large departure 
of $\Delta M_{B_s}$ from the value predicted in the SM.

Essential for identifying the sources of flavour violation are  
correlations among different observables. To give an example, suppose 
that significant deviations from the SM predictions are found in the
$B^0_d\rightarrow\phi K_S$ and $B^0_d\rightarrow\eta^\prime K_S$ decays. 
If they are due to LR/RL mass insertions, no deviation is expected in 
the $B^0_s$-$\bar B^0_s$ mixing. On the other hand, if the ``anomaly'' is due 
to LL/RR mass insertions, the $B^0_s$-$\bar B^0_s$ mixing should also be 
significantly affected. Thus, different SUGRA textures lead to different 
signatures. In principle, a more sophisticated network of correlations, 
including various other observables, can and should be developed. This, 
however, is beyond the scope of the present paper.

Significant SUSY effects can also be present in the top quark decays.
The magnitude of  $(\delta_{LR}^u)_{i3}$ mass insertions is enhanced 
due to a large top Yukawa coupling. Since BR($t\rightarrow c \gamma$)
is particularly sensitive to LR insertions,
BR($t\rightarrow c \gamma$)$\sim  10^{-6} (\delta_{LR}^u)_{23}^2 $ \cite{Delepine:2004hr},
large departures from the SM prediction $\sim 10^{-13}$ are expected.

To conclude, we see that the current $B-$physics
experiments are beginning to probe a natural range of supergravity non--universality
in the A--terms. Some processes can further probe non--universality in the
soft scalar masses, yet in this case it would be difficult to define 
a ``natural'' range due to larger model-dependence.
Correlations among various observables can allow one to identify the source
of flavour violation.

\section{Conclusions}

In this paper, we have presented a classification and analysis
of flavour violating sources in general supergravity models.
The current flavour physics data lead us to the following conclusions:
\begin{enumerate}
\item Flavour violation through the K\"ahler potential is disfavored,
      but room for flavour non--universality remains if the K\"ahler metric
      is diagonal. This often occurs in string models in which the K\"ahler 
      potential is protected by string selection rules.
\item Departures from  universality of order unity are allowed in the squark 
      sector $A-$terms. Such departures are expected in typical string 
      models.
\item $A-$terms in the charged slepton sector must be very well 
      aligned with the lepton Yukawa matrix, which points at either a 
      diagonal Yukawa matrix or  universal $A-$terms. 
\item A common and rather serious problem in supergravity models
      is the SUSY CP problem.
\item The FCNC problem  depends strongly  on the Yukawa texture
      and is  much milder for hierarchical Yukawa matrices.
      For diagonal squark masses, the problem essentially disappears 
      if the Yukawa matrices are diagonalized by small angle rotations or 
      if the CKM matrix  derives  entirely from rotations of the up type 
      quarks. Thus, there exist varieties of textures ensuring sufficient
      suppression of FCNC.
\item Importance of SUSY flavour effects grows with quark masses.
\item Current $B-$physics experiments are beginning to probe  a natural range
      of flavour non--universality in SUGRA models. Correlations among different
      observables can allow one to identify  the source of flavour 
      violation.
\end{enumerate}

\begin{table}[h]
\begin{center}
\begin{tabular}{|c||c|c|c|}
\hline
Texture & Hierarchical Yukawas & Democratic Yukawas & Mode \\
\hline
(A) & no problem & no problem  & ? \\
(B) & no problem & problem & $b\rightarrow s\gamma$, $b\rightarrow s\bar ss$\\
(C) & no serious problem  &  problem  & $B^0_s$-$\bar B^0_s$ \\
(D) & no serious problem  &  problem  & see (B), (C)  \\
(E) & problem            &  problem  & see (B), (C) \\
\hline
\end{tabular}
\end{center}
\caption{
Compatibility of the supergravity soft term textures with the Yukawa textures.
Possible detection modes are also indicated.
\label{table2}}
\end{table}

It is important to note the difference between the flavour structure of 
the Standard Model and that of supergravity soft terms. In the former, we encounter a 
hierarchical pattern of the Yukawa couplings. In supergravity, the soft 
terms are logarithmic derivatives of hierarchical quantities. As a result,  
$m^2_i$ and $\tilde A_{ij}$  are expected to be of the same order,
in sharp contrast with the Yukawa couplings.

Finally, from the point of view of avoiding the FCNC problem,
Ans\"atze (A) and (B) are clearly preferred. 
Ansatz (B) represents a rather typical prediction
of string models (e.g. heterotic string, intersecting branes, etc.).
Textures (C) and (D) are also allowed under the condition
of some degeneracy in the (12) block. Texture (E) is clearly disfavored.
These conclusions are summarized in Table 3.
The main message is that the supergravity textures are not necessarily
restricted to the universal one and can be quite rich, depending
on the mechanism generating the Yukawa matrices. These flavour structures
can be probed experimentally.
  
\vskip0.3cm

\noindent {\bf Acknowledgements}\\
This work has been supported in part by the RTN European Program 
MRTN-CT-2004-503369. P.H.Ch. and S.P. were supported by the Polish State 
Committee for Scientific Research Grants 
2 P03B 040 24 for 2003-2005 and 
2 P03B 129 24 for 2003-2005, respectively.
P.H.Ch. would like to thank the CERN Theory Group for hospitality during
the completion of this work.
This work was started during S.P.'s visit to the University of
Hamburg. The visit was possible owing to ``Forschung--Preis'' of the
A. von Humboldt Foundation. S.P. thanks Jan Louis for his
hospitality.
This collaboration was stimulated by an ENTApP 
sponsored visitor's programme on dark matter at CERN, 17 January - 4 
February 2005.

\newpage

\newpage

\newpage

\newpage


\newpage

\newpage

\newpage

\newpage

\newpage

\newpage


\newpage

\newpage

\newpage

\newpage

\newpage

\end{document}